\documentclass[a4paper,11pt]{article}
\pdfoutput=1 % if your are submitting a pdflatex (i.e. if you have
\usepackage{jinstpub} % for details on the use of the package, please
 
\usepackage{amsmath}
\usepackage{amssymb}
\usepackage{epsfig}
\graphicspath{{./figures/}}
\usepackage{subfig}
 \usepackage{placeins}

\title{The laser-based gain monitoring system of the calorimeters in the Muon  $g-2$ experiment at Fermilab}

\author{A.~Anastasi$^a$,
A.~Basti$^{a,c}$,
F.~Bedeschi$^a$,
A.~Boiano$^b$,
E.~Bottalico$^{a,c}$,
G.~Cantatore$^{d,e}$,
D.~Cauz$^{d,f}$,
A.T.~Chapelain$^g$,
G.~Corradi$^h$,
S.~Dabagov$^{h,i,j}$,
S.~Di~Falco$^a$,
P.~Di~Meo$^b$,    
G.~Di~Sciascio$^k$, 
R.~Di~Stefano$^{b,l}$,
S.~Donati$^{a,c}$,
A.~Driutti$^{d,f}$,
C.~Ferrari$^{a,m}$,  
A.T.~Fienberg$^{n}$,
A.~Fioretti$^{a,m,1}$,
C.~Gabbanini$^{a,m}$,
L.K.~Gibbons$^{g}$,    
A.~Gioiosa$^{k,o}$,  
P.~Girotti$^{a,c}$,  
D.~Hampai$^h$,
J.B.~Hempstead$^{n}$, 
D.W.~Hertzog$^{n}$,
M.~Iacovacci$^{b,q}$,
M.~Incagli$^a$,
M.~Karuza$^{d,q}$,
J.~Kaspar$^{n}$,
K.S.~Khaw$^{n}$,
A.~Lusiani$^{a,r}$,
F.~Marignetti$^{b,l}$,
S.~Mastroianni$^b$,   
S.~Miozzi$^k$,
A.~Nath$^b$,    
G.~Pauletta$^{d,f}$,
G.M.~Piacentino$^{k,o}$,
N.~Raha$^a$, 
L.~Santi$^{d,f}$,
M.~Smith$^{a,n}$,  
M.~Sorbara$^{k,s}$,  
D.A.~Sweigart$^{g}$ and   
G.~Venanzoni$^{a,1}$\note{Corresponding author}}
\affiliation[a]{INFN, Sezione di Pisa, Largo Bruno Pontecorvo 3, I-56127 Pisa, Italy}
\affiliation[b]{INFN, Sezione di Napoli, Complesso Universitario di Monte Sant'Angelo, via Cinthia, I-80126 Napoli, Italy}
\affiliation[c]{Dipartimento di Fisica, Universit\`a di Pisa, Largo Bruno Pontecorvo 3, I-56127 Pisa, Italy}
\affiliation[d]{INFN, Sezione di Trieste, via A. Valerio 2,  I-34127 Trieste, Italy}
\affiliation[e]{Dipartimento di Fisica, Universit\`a di Trieste, via A. Valerio 2, I-34127  Trieste, Italy}
\affiliation[f]{INFN, G.C. di Udine e Universit\`a degli Studi di Udine, Via delle Scienze 208, I-33100 Udine, Italy}
\affiliation[g]{Cornell University, Department of Physics, 109 Clark Hall, Ithaca, NY 14853, USA}
\affiliation[h]{Laboratori Nazionali di Frascati dell'INFN,  Via Enrico Fermi 40, I-00044 Frascati, Italy}
\affiliation[i]{P.N. Lebedev Physical Institute RAS, Leninsky Ave. 53, 119991 Moscow, Russia}
\affiliation[j]{National Research Nuclear University "MEPhI", Kashirskoe Str. 31, 115409 Moscow, Russia}
\affiliation[k]{INFN, Sezione di Roma Tor Vergata, via della Ricerca Scientifica 1, I-00133 Roma, Italy}
\affiliation[l]{Universit\`a di Cassino, Via G. Di Biasio 43, I-03043 Cassino, Italy}
\affiliation[m]{Istituto Nazionale di Ottica, CNR-INO, S.S. 'A Gozzini',
  via G. Moruzzi 1, I-56124 Pisa, Italy}
\affiliation[n]{University of Washington, Department of Physics, Box 351560, Seattle, WA, 98195 USA}
\affiliation[o]{Universit\`a degli studi del Molise, Dipartimento Bioscienze e Territorio, Contrada Fonte Lappone, I-86090 Pesche, Italy}
\affiliation[p]{Dipartimento di Fisica,  Universit\`a di Napoli "Federico II", Complesso Universitario di Monte Sant'Angelo, via Cinthia, I-80126 Napoli, Italy}
\affiliation[q]{Department of Physics and Centre for Micro Nano Sciences and Technologies, University of Rijeka, Radmile Matejcic 2, 51000 Rijeka, Croatia}
\affiliation[r]{Scuola Normale Superiore, Piazza dei Cavalieri 7, I-56126 Pisa, Italy}
\affiliation[s]{Dipartimento di Fisica dell'Universit\`a di Roma Tor Vergata, via della Ricerca Scientifica 1, I-00133 Roma, Italy}

\emailAdd{graziano.venanzoni@pi.infn.it, andrea.fioretti@ino.cnr.it}

\abstract{The Muon $g-2$ experiment, E989, is currently taking data at Fermilab with the aim of reducing the experimental error on the muon anomaly by a factor of four and possibly clarifying the current 
discrepancy with the theoretical prediction. A central component of this four-fold improvement in precision is the  laser calibration system of the calorimeters, which has to monitor the gain variations of the photo-sensors with a 0.04\% precision on the short-term ($\sim 1\,$ms). This is about one order of magnitude better than what has ever been achieved for the calibration of a 
particle physics calorimeter. The system is designed to monitor also long-term gain variations,
 mostly due to temperature effects, with a precision below the per mille level. This article reviews the design, the implementation and the performance of the Muon $g-2$ laser calibration system, showing how the experimental 
requirements have been met.}


\begin{document}
\maketitle
\flushbottom

\section{Introduction}
\label{sec:intro}
%Physics motivations, status of the whole experiments (Franco Bedeschi, Pisa)

As the muon anomaly $a_{\mu}=(g-2)/2$ can be computed to an extremely high precision~\cite{Jegerlehner2008, Blum2013}, the new Muon $g-2$ experiment at Fermilab, E989~\cite{Carey2009, Grange2015, Hertzog2016}, will provide an important test of the Standard Model, and a sensitive search for new physics~\cite{Stokinger2009}.

The E989 experiment aims to improve the precision on the measurement of the muon anomaly by about a factor of four relative to the previous Brookhaven experiment, E821~\cite{Bennett2004, Bennett2006}, i.e. to an uncertainty of $16 \times 10^{-11}$ ($\pm 0.14$~ppm). A large fraction of this improvement is obtained by increasing the total number of muon decays recorded, but then the systematic uncertainty has to match a target statistical error of 0.1~ppm. This requires a control of the systematics at the level of 2 to 3 times better than previously achieved, in particular at about 0.07~ppm on the knowledge of the anomalous precession frequency of the spin of the muon ($\omega_a$) and on the precession frequency of protons at rest ($\omega_p$), which is used to measure the magnetic field. To achieve the statistical uncertainty of $0.1$~ppm, the total data set must contain approximately $1.5 \times 10^{11}$ detected positrons with energy greater than $1.5$~GeV. 

The experiment consists in filling  a storage ring with polarized muons and measuring the anomalous precession frequency $\omega_a$. The latter is achieved by measuring the modulation of the rate of  
positrons produced by muon decays. The positrons are detected by 24 
calorimeter stations located around the storage ring. Muons are injected in the storage ring in bunches; their decays are 
observed  for $\sim 700$~$\mu$s ({\it fill} time) after the injection, then the 
few remaining muons are discarded and a new bunch is injected into the ring.

%The muon  {\it fill} lasts only $\sim 700$~$\mu$s after which most of the muons have decayed. The remaining muons are then discarded and  a new fill is injected into the ring.

Because of the increased rate of muon decays and the more demanding goals in systematic uncertainties,
the collaboration had to devise improved instrumentation~\cite{Grange2015} with respect to the E821 experiment. 
%The kickers that insert the muons in the correct orbit are improved.
%%in the storage
%%ring are entirely new and optimized to give a precise kick on the first turn only, increasing the stored
%%fraction. 
%The magnetic field is more carefully prepared and monitored. The detectors and
%electronics are entirely new. New in-situ trackers  provide precise information on the stored beam.
In particular, a set of 24 calorimeters, with matrices of 6x9 PbF$_2$ crystals, equipped with fast, non-magnetic photo-detectors, are placed around the ring, and provide large acceptance of decay positrons with precise energy and time reconstruction. A precision laser calibration system ensures critical performance
stability of these detectors throughout the long data taking periods, as well as the synchronization of different detectors and the debugging of the whole data acquisition system.

The PbF$_2$ crystals (dimensions $25 \times 25 \times 140 \, {\rm mm}^3$) produce Cherenkov light when hit by a muon-decay positron~\cite{Savchenko2017}. This light is detected by large-area Silicon Photo-Multipliers, SiPMs, which turn the light into an electric signal that is digitized in 1.25~ns bins by custom 
electronics~\cite{Grange2015}. During the fill, these sensors are subject to large changes in the rate of particles leaving signals in the calorimeters. Indeed in the initial phase there is a large muon intensity as well as a flash of additional quickly decaying background. After a few $\mu$s the background has disappeared and the muon rate decreases exponentially with a time constant of 64~$\mu$s. These large rate variations affect the SiPM response due to both the intrinsic sensor performance and small variations in the sensor voltage provided by the power supplies~\cite{Kaspar2017}. 

One of the most relevant systematic errors of the experiment originates from variations of the gain of the calorimeters. They affect the measurement of $\omega_a$ very significantly, especially as they occur also during the time scale of the muon fill. 
%In order to reach the desired sensitivity, the experiment must be able to correct the average gain variation during the fill with an accuracy at the $4\times 10^{-4}$  level, or 0.04\%. 
Additional slower variations of the gain due to environmental effects on the light sensors, such as the temperature, must also be monitored and corrected for, in order to allow the determination of an absolute energy scale and therefore a precise interpretation of all the quantities determined in the analysis of the data.

This paper describes the laser calibration system that is used to monitor the gain fluctuations,  to provide the calibration constants and  the time reference to the calorimeters, and to emulate the time distribution of the signals coming from the muon decays. It is organized as follows: Section~\ref{sec:distribution} describes the complete laser distribution system, which for clarity has been decomposed into the diffusing system,~\ref{ssec:diffuser}, the front panel,~\ref{ssec:panel}, the double-pulse setup,~\ref{ssec:doublepulse}, and other detectors,~\ref{ssec:other}. Section~\ref{sec:monitors} describes the two systems devoted to monitor the stability of the laser source,~\ref{ssec:source}, and of the distribution chain,~\ref{ssec:local}. Section~\ref{sec:electronics} describes the control electronics,~\ref{ssec:control}, and the readout system,~\ref{ssec:processing}. Section~\ref{sec:operation} describes the system performance in different laser operation modes: standard operation mode,~\ref{ssec:standard}, double-pulse gain calibration,~\ref{ssec:double}, absolute energy calibration,~\ref{ssec:absolute}, and flight simulator mode,~\ref{ssec:flight}.

\section{The laser distribution system}
\label{sec:distribution} %(AF, CG, CF Pisa)

\begin{figure}[htbp]
\centering
\includegraphics[width=0.85\textwidth]{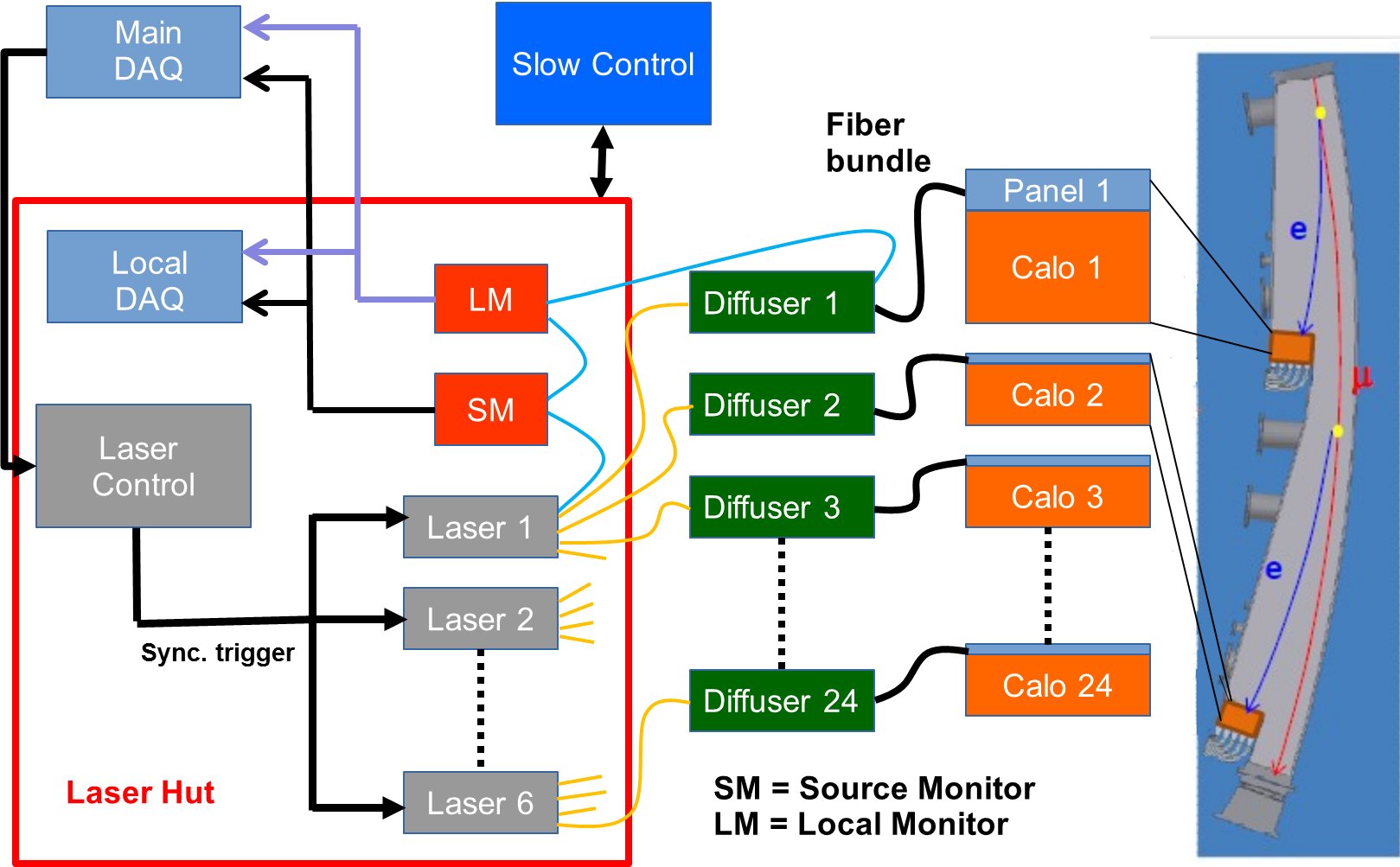}
\caption{Schematic of the laser distribution system. For clarity each laser box indicates both the Laser Head and the optical setup to split the laser beam in four part to be sent to 4 calorimeters. Also for clarity the links that exist between each Laser Head and its Source Monitor and between each Diffuser and its Local Monitor are shown only for one SM and LM respectively.}
\label{fig:distribution}
\end{figure}

The E989 experiment requires a continuous monitoring and calibration of the calorimeters photo-detectors at the 0.04\% level on the short timescale of a single beam fill ($700\,\mu$s). Monitoring and correction of the gain over the longer term of an entire run 
(several hours) below the per mille level is also required.
%whose
%response may vary on both the short timescale of a single beam fill (700~$\mu$s), and the long one of the duration of each single run which lasts up to a few hours.
%In order to achieve the design systematic uncertainties, it is estimated that the detector response must be
%calibrated with a relative accuracy at sub-per-mille level on the short term scale of the beam fill
%and at the sub-per-cent scale over the longer term of the entire data collection. 
This is a challenge for the
design of the calibration system because the desired precision is at least one order of magnitude higher
than that of all other existing or past calibration systems for calorimetry in particle physics~\cite{Rogan2010,vanWoerden2016,Zhang2019}.
The proposed solution, whose scheme is sketched in Fig.~\ref{fig:distribution}, is based on the method of sending simultaneous light calibration pulses onto each of
the 1296~crystals of the electromagnetic calorimeter. Light pulses must be stable in intensity and
timing in order to correct drifts in the response of the crystal readout devices. The stability of the laser
intensity is monitored with a suitable photo-detector system.

The filling of the muon ring by the Fermilab beamline occurs with a particular time sequence (see Fig.~\ref{fig:Pulse_generation}), 
where the main clock cycle has a period of 1.4
seconds. At each cycle 2 groups of 8 bunches each are sent to the storage
ring with a frequency of 100~Hz. Each bunch is observed in the ring for
approximately 700~$\mu$s fill time,     
corresponding to roughly 11 muon lifetimes. Laser calibration pulses can be sent, according to a strategy, described in Sec.~\ref{ssec:standard} and illustrated in Fig.~\ref{fig:Pulse_generation}, before or after each fill (called Out-of-Fill Pulses, OoFP) or inside a fill (called In-Fill Pulses, IFP).

As the laser calibration pulses must arrive simultaneously to all the channels located in 24 calorimeters around the 14-meters diameter Fermilab muon storage ring, we chose to put most of the laser apparatus in a dedicated room, the Laser Hut, located inside the $g-2$ experiment hall. It is a 4 by 4-meters wide, light-tight, acoustically isolated and thermally controlled room, from where laser calibration pulses are sent to calorimeters through multimode optical fibers. 
%All optical control signals are thus recorded only inside the Laser Hut. 
Additional laser signals for timing/calibration purposes are also sent from the Laser Hut to two other $g-2$ detectors called $T0$ and Fiber Harps. All monitor signals of the laser power are recorded only inside the Laser Hut, thus avoiding any electric contamination from signals coming from other detectors. 
The crucial elements for the realization of this system are: 1) the light source; 2) the
distribution system that distributes the light to the calorimeters with adequate intensity and homogeneity, and 3) the monitoring system.
The light wavelength must be in the range of the calorimeter photo-detector sensitivity and the light source
must have an adequate power to deliver an appropriate amount of light to all crystals. 

\begin{figure}[htbp]
\centering
\includegraphics[width=0.85\textwidth]{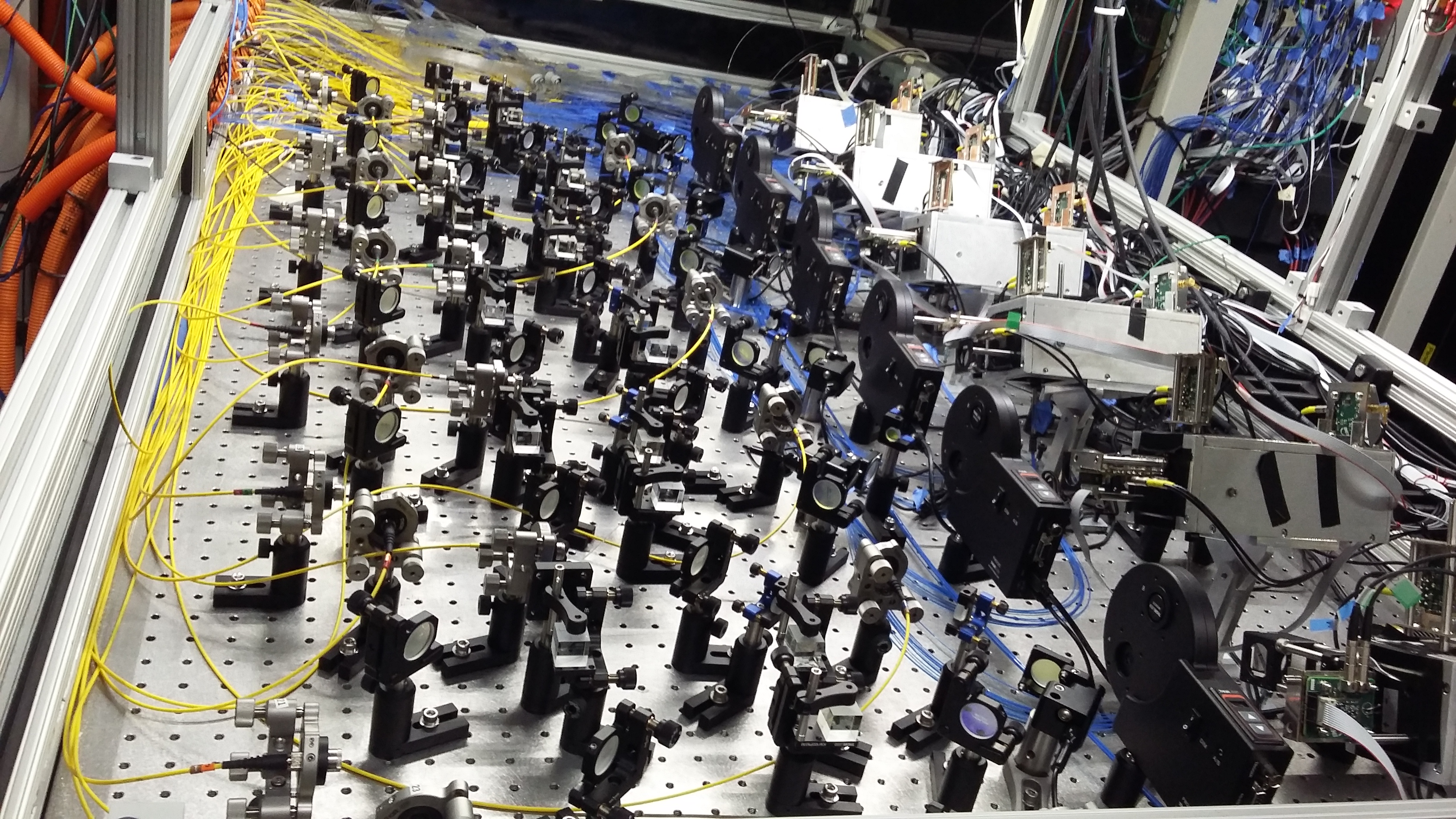}
\caption{Picture overviewing the optical table. On the right side are the laser heads, partially hidden by the
Source Monitors (silver aluminum boxes). The laser beams are then attenuated by 6 motorized filter wheels (black aluminum). In the center the optical elements used to split each laser beam in four parts, each one injecting a launching fiber (yellow cables). On the far right, the rack with the Local Monitor boxes, with the reference signals coming from the Source Monitors through optical fibers (blue cable). On the far left the orange plastic tubes guiding launching and monitoring optical fibers to the 24 calorimeters. Zoom in for a more detailed view of the components.}
\label{fig:table}
\end{figure}

Given these requirements, 
the chosen setup is the following (see scheme in Fig.~\ref{fig:distribution}, and picture in Fig.~\ref{fig:table}).
Laser calibration pulses for the 24 calorimeters are generated by 6 identical laser diode heads (Picoquant, mod.~LDH-P-C-405M), 700~pJ/pulse, 600~ps (FWHM) duration, each one serving 4 calorimeters. This has been a somewhat conservative choice, as the laser power available exceeds what is actually required by over a factor 4, but it allows special calibration modes, i.e. the double pulse mode (see Sec.~\ref{ssec:doublepulse} and~\ref{ssec:double}), and recovery, in case of failure of one or more laser heads. The laser heads are driven by a multi-head controller (Picoquant, mod. PDL 828 "Sepia II") and can be separately triggered.  The power of each laser head can be varied either in discrete steps in a reproducible way by remotely changing neutral density filters through a filter wheel (Thorlabs, mod.~FW212CWNEB), or continuously by varying the current of the driver.
Laser signals to the 24 calorimeters are sent through 24 silica {\it launching} optical fibers (Polymicro, mod.~FDP400440480) 25-meters long, 400~$\mu$m diameter. while laser monitoring signals from calorimeters are sent back to the Hut by 48 fibers: 24, 1~mm diameter PMMA fibers (Mitsubishi Eska, mod.~GK40)  and 24, 600~$\mu$m diameter silica fibers (Thorlabs, mod.~FP600URT).
A sketch of the laser distribution system is shown in Fig.~\ref{fig:distribution}.  Splitting of each laser output into four beams and subsequent injection into the launching fibers is performed in air through standard optical elements placed upon a temperature controlled, light-tight optical table. A fiber coupler (Thorlabs, CFC-8X-A) is placed at the entrance of each launching fiber. Roughly 70\% of the incoming light is coupled into each fiber. The launching fiber and the two monitoring fibers of each calorimeter are inserted into protective, semi-rigid
plastic tubes that guide them inside the ring to each calorimeter, keeping them separated from the other electric cables. 

\subsection{The diffusing system}
\label{ssec:diffuser} 

\begin{figure}[htbp]
\centering
\includegraphics[width=0.75\textwidth]{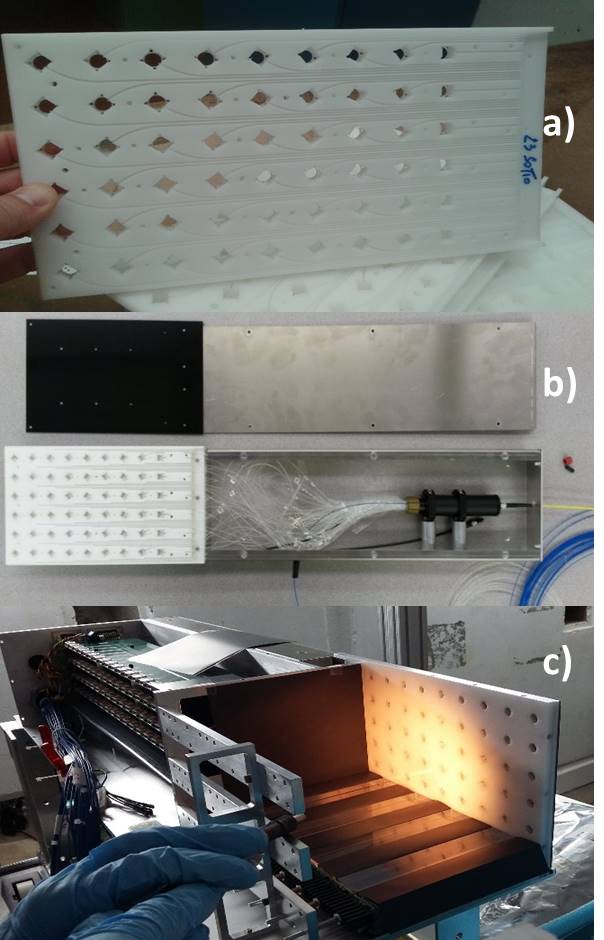}
\caption{Picture showing: (a) the Delrin front panel, (b) the diffusing box, with the diffuser and the fiber bundle, connected to the Delrin panel and the optical prisms, and its light-tight cover, (c) the assembling of the PbF$_2$ crystals on the calorimeter equipped with the front panel.}
\label{fig:panelo}
\end{figure}

In order to send the light pulses to all 54 detectors of each of the 24 calorimeter, each launching fiber is connected to a dedicated box (see Fig.~\ref{fig:panelo}b) close to the calorimeter where the light output is collimated and transmitted through an engineered diffuser  (Thorlabs, mod.~ED1-S20), consisting of structured microlens arrays that transform a Gaussian input beam into a flat top one~\cite{Anastasi2015}. A fiber bundle made of 60 (54 plus spares) 1~mm-diameter PMMA fibers is positioned a few cm from the diffuser and uniformly illuminated by the light from the diffuser. On the external part of the bundle ferrule there are some slots where the two monitoring fibers (in quartz and in PMMA) are inserted. They are dedicated to the so-called Local Monitor (LM), and go back to the Laser Hut where the LM is placed. All optical elements forming the diffusing box, i.e. the collimating lens, the diffuser and the fiber bundle, are mounted on a single, light-tight, 1-inch lens tube. As the light beam transmitted through the diffuser diverges (by 20~degrees), the light collected by the fiber bundle depends significantly on the distance set inside the tube between the diffuser and the bundle. 

We performed a detailed study on the homogeneity of the light profile as a function of the diffuser-bundle distance~\cite{Anastasi2015}, concluding that for distances larger than 40~mm the light inhomogeneity becomes of the order of few percent, i.e. lower than other factors determining the final light intensity distribution delivered to the 54~crystals, which include fiber surface machining and fiber to panel coupling (see Sec.~\ref{ssec:panel}). In this configuration, the light collected and transmitted by each 1~mm fiber of the bundle is of the order of some $10^{-4}$ of the total light impinging on the diffuser, which is sufficient for our purposes. We also studied the temporal stability of this element of the distribution system by monitoring the ratio of the light transmitted by two different fibers of the bundle~\cite{Anastasi2015}. The temporal stability exceeds our requirements, indicating that processes potentially unsafe for a stable pattern, like speckle formation, are negligible.

\subsection{The front panel}
\label{ssec:panel} 

In the experiment,  positrons emitted by muon decay curl towards the interior of the ring, where they can be collected by one of the 24 electromagnetic calorimeters located at regular intervals along the inner perimeter of the storage ring (see far right side of Fig.~\ref{fig:distribution}). The calorimeter is segmented into six rows and nine columns of lead-fluoride (PbF$_2$) crystals~\cite{Fienberg2015}. Each crystal is read out by a SiPM photo-detector whose gain shows substantial variations over time~\cite{Kaspar2017} so that it has to be continuously calibrated.

\begin{table}[htbp]
	\centering
		\begin{tabular}{|l|l|l|}
		\hline
		  Position in 										& Energy per		&		notes	  		\\
			distribution/monitoring chain		& laser pulse		&				    		\\	
		\hline
		Laser  head										& $700\,$pJ			&   \\
		After Source Monitor					& $490\,$pJ			&   \\
		Onto Source Monitor						& $210\,$pJ			&		\\
		Towards Fiber Harps detectors	&	$70\,$pJ			&		Only from one laser head\\
		After filter wheel						& $170\,$pJ			&		From now on values are with filter n. 6\\
																	&								&		(standard working mode) \\
	  Before each launching fiber		&	$40\,$pJ			&		\\
		Onto each diffuser						& $30\,$pJ			&	 	\\
%		Onto each bundle							& $3\,$pJ				&	 	\\
		Onto each front panel's fiber	& $30\,$fJ			&	Diffuser-bundle at $40\,$mm 	\\
																	&								&	working distance	\\
		Onto each PbF$_2$ crystal			&	$18\,$fJ			&		\\
		Onto each SiPM								& $4\,$fJ				&		\\
		Onto LM from diffuser					&	$4-6\,$fJ			&		\\
		Onto T0 detector							&	$4-6\,$fJ			&		\\
		Onto Fiber Harps 							&	$5-10\,$fJ		&		\\
\hline
		\end{tabular}
\caption{Laser energy per single pulse at various points of the laser distribution and monitoring systems. These values are only indicative and may vary by 5-10\% at each step going from one distribution line to another.}
\label{tab:LaserIntensity}
\end{table}

The vacuum chamber is specially shaped in correspondence with the calorimeters to allow positrons to travel mostly in vacuum until their final detection in the calorimeter. This configuration leaves very little space between the wall of the vacuum chamber and the front of the calorimeter (about 20~mm). This space was used to introduce a panel that collects the 54 fibers from the diffuser and conveys their output, by bending the light pulses by $90^\circ$, into each crystal.
The calibration laser pulses delivered by the optical fibers require an optical component to reach the surface of the crystals at normal incidence, as it is not possible to bend the fibers in the limited space available.
To this end, a thin panel was conceived (see Fig.~\ref{fig:panelo}). It is made of Delrin, a material characterized by its large radiation length and having appropriate mechanical properties. The panel, 12~mm thick, corresponding to a radiation length of $0.044{\rm X}_0$, holds the optical fibers in a fixed position with respect to the crystals. Inside the panel, each fiber is terminated onto a 10~mm side right-angle prism  that performs the final $90^\circ$ turn of the laser light into the crystals. 24 Delrin panels were made, each holding 54 prisms and fibers.
Fibers and prisms are held in place with glue; no optical grease has been inserted between fibers and prisms in order to avoid light transmission changes during the years due to aging of the grease. Finally, each panel is fixed to an aluminum box (see Fig.~\ref{fig:panelo}), containing the diffusing system, rigidly connected to the calorimeter assembly. Fig.~\ref{fig:panelo}c  shows the Delrin panel and the crystals during the assembling of the first prototype.

The laser energy per single pulse impinging in any part of the laser distribution and monitoring systems, in standard working mode, is reported for clarity in Table~\ref{tab:LaserIntensity}.

%
%\begin{table}
	%\centering
		%\begin{tabular}[b]			
		%\end{tabular}
	%\caption{Laser iintensity at various points of the laser distribution and monitoring systems}
	%\label{tab:LaserIintensity}
%\end{table}
%
\subsection{The double-pulse setup}
\label{ssec:doublepulse}

The Double Pulse setup allows a working mode in which 2 consecutive pulses
from two different lasers are sent to each crystal, with a 
delay which can vary from 1~ns up to $100 \,\mu$s.

\begin{figure}[htbp]
\centering
\includegraphics[width=0.48\textwidth]{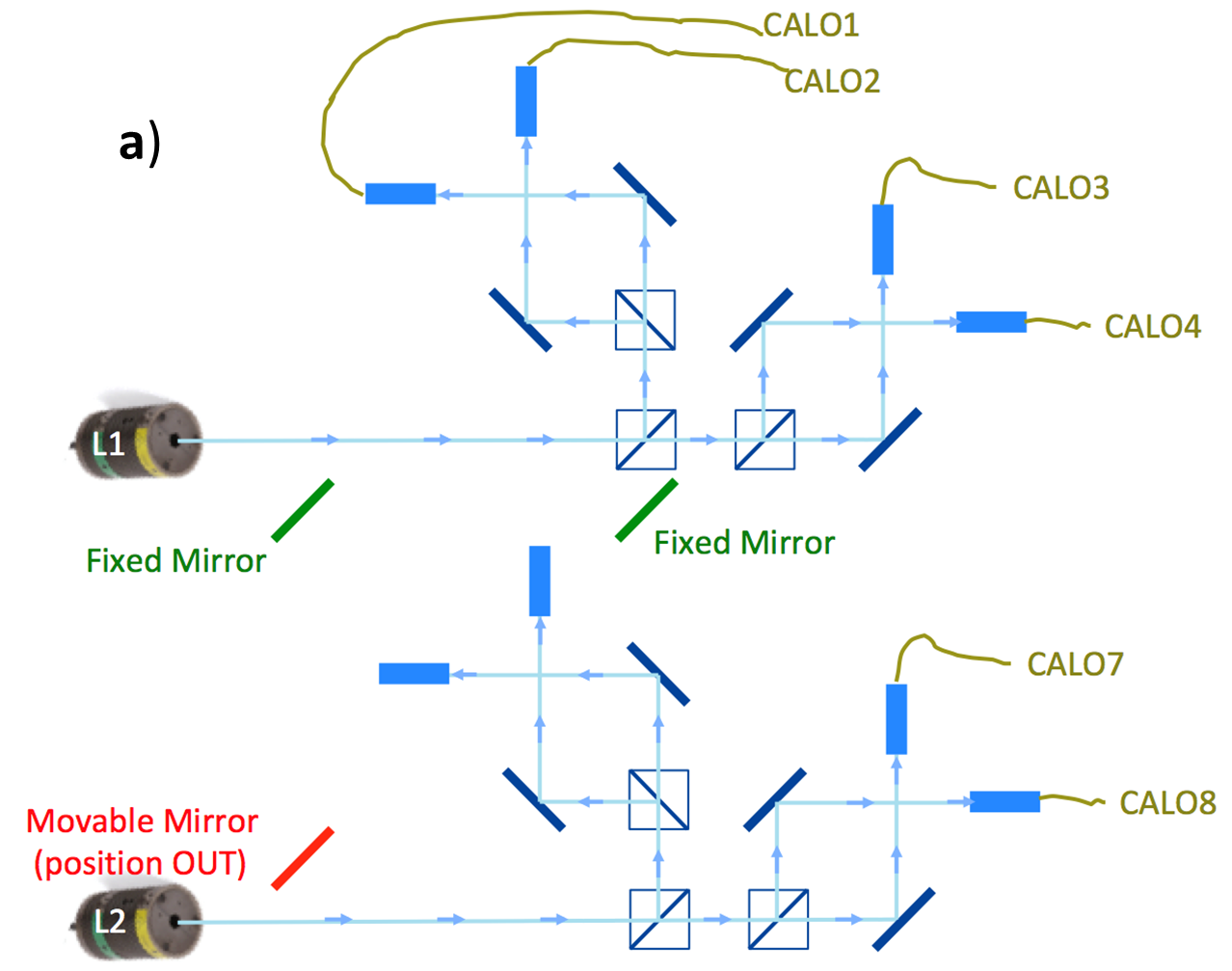}
\includegraphics[width=0.48\textwidth]{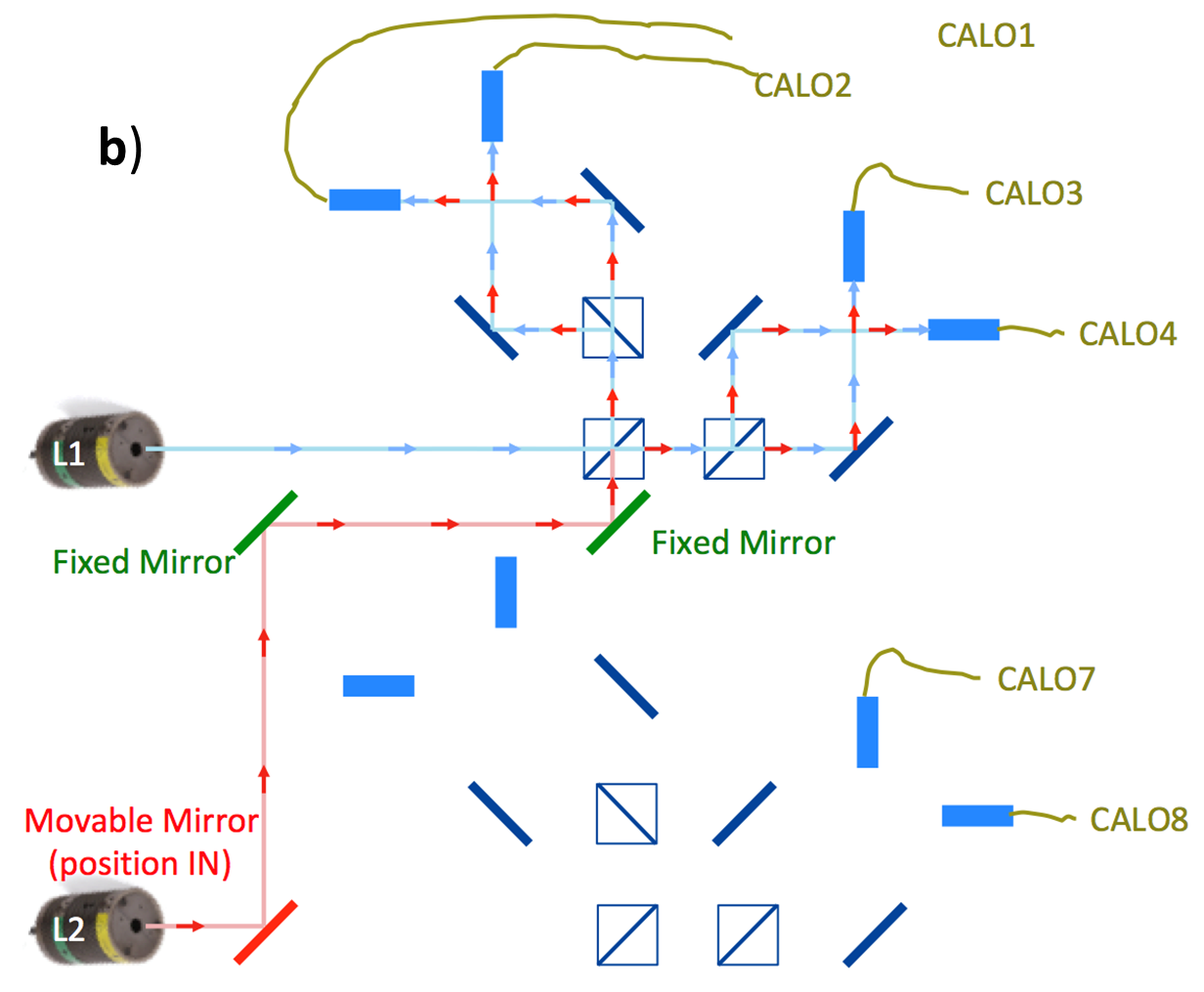}
\caption{Left: sketch of optical setup for the standard laser operation (movable mirror OUT) in which each laser is split in four and injected into 4 fibers. Right: modified setup for double pulse operation (movable mirror IN) in which one laser beam is superimposed into the path of its paired one. A similar set of moving+fixed mirrors (not shown for clarity) is present also on the upper laser setup to allow the symmetrical operation.}
\label{fig:optics1}
\end{figure}

In standard data taking, each laser sends light pulses to four
calorimeters, as described in Sec.~\ref{sec:distribution} and shown in Fig.~\ref{fig:optics1} a.
%The light is split in two, and then in four, by three {\em beam-splitter cubes} (BSC), as shown in Fig.~\ref{fig:optics1}a, and then directed to 4~quartz {\it launching} fibers by a mirror. A fiber coupler (Thorlabs, CFC-8X-A) is placed at the entrance of each fiber. Roughly 70\% of the incoming light is coupled into each fiber.  
By placing a movable mirror in front of each laser head, and the use of additional fixed mirrors, it is possible to form 3 pairs of lasers, 1-2, 3-4 and 5-6, coupled together. By varying the position of the movable mirrors, it is possible to re-direct the light of each laser into the path of its paired one through the first beam-splitter cube. The two laser beams are thus superimposed and injected, with comparable intensity, into the same fiber, as shown in Fig.~\ref{fig:optics1} b. In this way, one can remotely decide to send two laser pulses, with an adjustable time separation, into the same fiber, i.e. into 4 calorimeters. Of course, when doing this, the other 4 calorimeters, corresponding to the laser with the modified light path, do not receive any light. 

%\begin{figure}[htbp]
%\centering
%\includegraphics[width=0.75\textwidth]{optics2.png}
%\caption{...}
%\label{fig:optics2}
%\end{figure}

%In a similar way, the light of laser 1 can be re-directed into the
%light path of laser 2; the same happen to the other pairs: 3-4 and 5-6.

The movable mirrors are mounted on motorized standard {\em flip-flop}
%stands (Thorlab MFF101/M), remotely controlled through 6~USB to TTL-232RG cables 
%by a computer located in the Laser Hut. A C++
%program, that can be run remotely, operates the stands through the USB port. The {\em flip-flop} 
units (Thorlab MFF101/M) that allow for a
reproducibility of the beam alignment better than 0.1~mrad, which is enough for our purposes.

%\subsubsection{The Delay Generator}

\begin{figure}[htbp]
\centering
\includegraphics[width=0.75\textwidth]{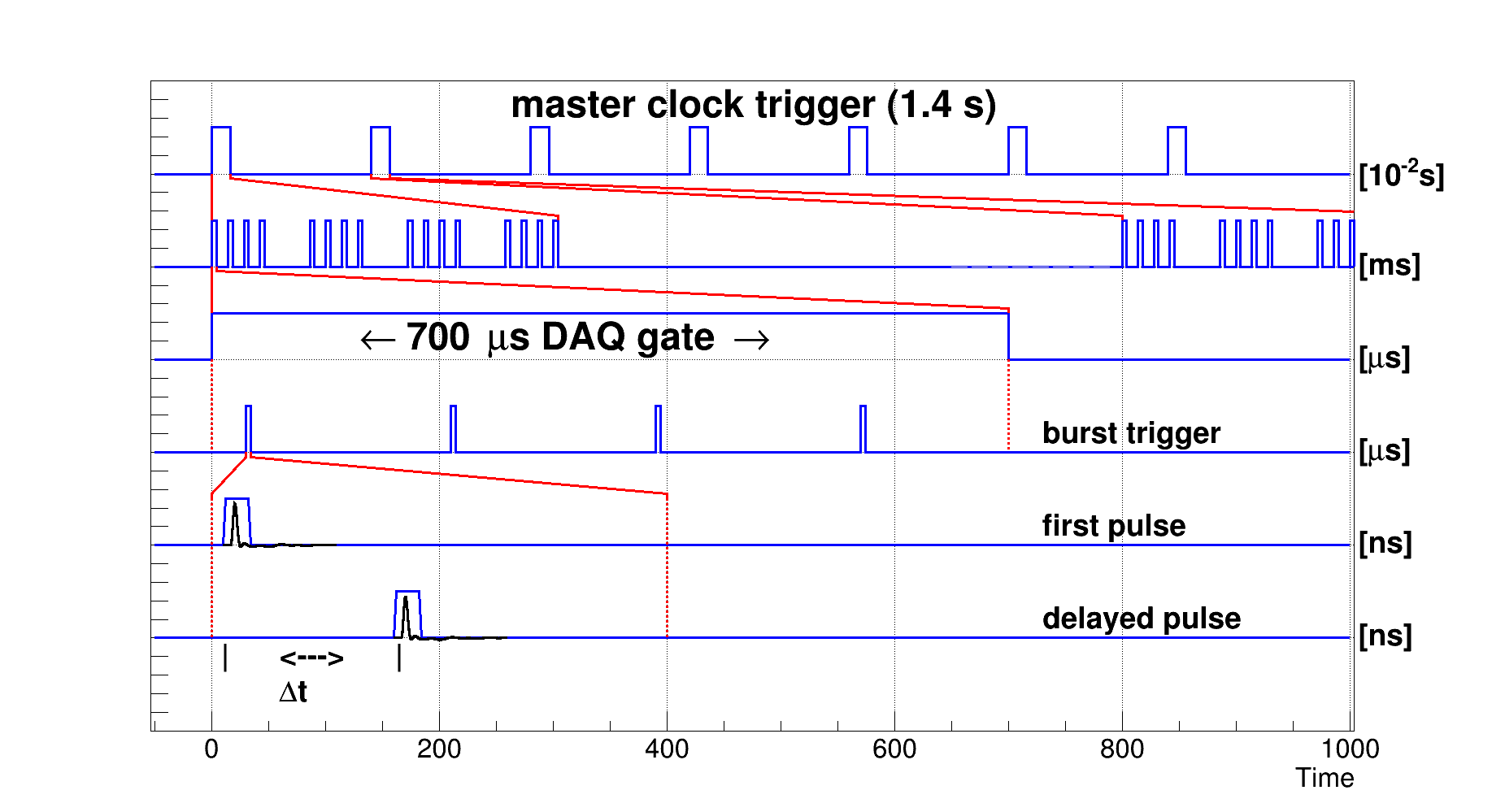}
\caption{Example of trigger structure for Double Pulse operation. A set of 4 double pulses, delayed by 10~ns, is produced within each muon fill. Blue squared pulses are electronic triggers, while black spikes represent laser pulses.}
\label{fig:DGtrigger}
\end{figure}

An external delay generator (SRS DG645) is used to send prompt and
delayed signals to the laser trigger. 
An example of how the pulses are triggered is shown in
Fig.~\ref{fig:DGtrigger}. 
%As the clock cycle has a period of 1.4 seconds and for each cycle 16 different bunches are sent to the storage ring with a frequency of 100~Hz. Each bunch stays in the ring for approximately 700~$\mu$sec, the so-called {\em fill time} corresponding to $~10$ muon life times.
Within each fill, a programmable set of triggers (a so-called {\em burst of
  triggers}) can be sent by the delay generator. Each trigger
corresponds to 2 output channels whose relative delay
can be programmed.
%~\footnote{The device allows to program up to 4
  %channels independently, and up to 8 with an extension placed on the
  %back. We are using just two of them for the Double Pulse task.}.
Fig.~\ref{fig:DGtrigger} shows a possible setup with a burst of 4~triggers within the fill. 
Each trigger provides 2~NIM pulses delayed by $10\,$ns 
(note the scale multiplier on the right), $50\,$ns long, used to trigger the two laser drivers and produce the two laser pulses.

\subsection{Time synchronization}
\label{ssec:other}

In addition to serving as a calibration system for the calorimeters, the laser provides a synchronization time signal to the calorimeters, the $T0$ counter, and the Fiber Harps.

The $T0$ counter is a $1\,$mm-thick scintillator panel with two PMTs on the same axis. It requires two laser pulses in order to synchronize the traces obtained from the PMTs. To this purpose, two 10~m long, $1\,$mm diameter, PMMA optical fibers have been connected by SMA mating sleeves to two spare fibers coming from the diffusers of two calorimeters.

The Fiber Harps, two beam profile detectors that can be inserted in the muon path in two positions around the ring, are more demanding. Each detector has two pairs of 8 small SiPMs arrays (for X and Y profiles), each SiPM collecting the signals from a scintillating fiber. Additional 0.5~mm PMMA, optical fibers have been connected to the SiPMs in order to receive laser calibration pulses. The 16 PMMA fibers of each detector are grouped in one bundle, fed by a launching fiber in the very same way as the 24 calorimeters (see Section~\ref{sec:distribution}) for synchronization and calibration purposes. The collimators of the two launching optical fibers, connected to two engineered diffusers on the detector side, are illuminated with a laser beam sampled from one of the laser heads by a 10:90 (R:T) ratio non-polarizing beam-splitter cube (Thorlabs, mod.~BS025) located before the filter wheel. The amount of light, delivered to the Fiber Harps bundles by this optical arrangement, is similar to that delivered to each calorimeter and can be reduced by an additional filter wheel placed after the cube.

\section{The laser monitoring system}
\label{sec:monitors}
The stability of the power delivered by the lasers to the calorimeters is monitored by two different systems: 
a set of six Source Monitors (SM), each one placed directly after each laser head, and a set of 48 Local Monitors, two of them after each diffusing system.  

\subsection{The Source Monitor}
\label{ssec:source}
%(Giovanni Pauletta, Udine)
%The source monitor system  measures the average laser power using two PIN diode
%photodetectors and a photomultiplier (PMT) that simultaneously also measures the energy released by a
%$^{241}{\rm Am}$ source in a NaI crystal. The Picoquant lasers heads are very stable (1\% RMS over 12~hours and 3\%
%peak-to-peak for $\Delta T_{\rm amb} < 3$~K), and the SM is able to monitor laser power changes at the per-mil level.

The Source Monitor monitors the stability of the laser source. The SM was designed to satisfy statistical requirements rapidly while minimizing systematics such as those arising from the gain variation of the photo-detector or temperature instabilities. To this purpose, 30\% of the total laser intensity is directed into the SM by a beam splitter located just downstream of the laser source. The design allows for correction of shot-to-shot fluctuations at the per-mille level as well as variations in the average intensity at the required 
(0.04\%) precision in about 100 shots. Sensitivity to fluctuations due to mechanical vibrations, intrinsic response, external electronic noise and to variations in beam pointing are also minimized. A reference light source is also incorporated, allowing for long-term monitoring of the average intensity.  

\begin{figure}[htbp]
\centering
\includegraphics[width=0.75\textwidth]{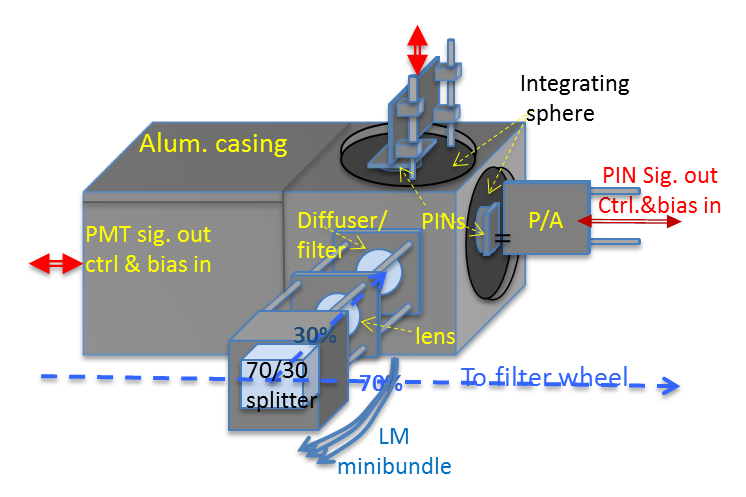}
\includegraphics[width=0.75\textwidth]{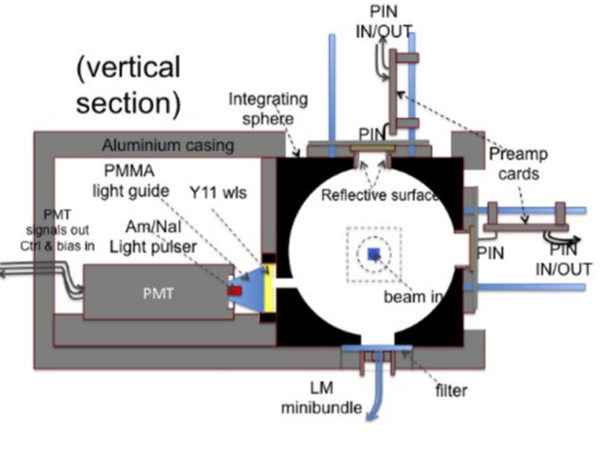}
\caption{(a) External view of  Source Monitor. The laser beam is sampled directly after the laser head by a 30/70 beam splitter cube, and directed into an integrating sphere after a lens and a diffuser. (b) Vertical Section of SM. Two large-surface PIN diodes are coupled directly to two ports of the integrating sphere, while a third one is used to send light to a PMT and to an optical fiber for referencing of the Local Monitor. A $^{241}$Am/NaI light pulser for absolute light reference is positioned in front of the PMT.}
\label{fig:SM}
\end{figure}

%Two alternative designs of the SM were developed and tested. Both are characterized by a rigid mechanical structure, with a large thermal inertia and good electrical shielding which contains all detector components. These designs differ with respect to the method employed to eliminate beam-pointing fluctuations. The first design combines an engineered diffuser with a reflective chamber as already realized for the  diffusing system in the laser distribution system while, in the second case, this combination is substituted by an integrating sphere.  Both were tested and resulted to satisfy requirements~\cite{Anastasi2017}.  As the second design can be more easily integrated with the LM  (see Section~\ref{ssec:local}), it was preferred over the first one.  The design of the adopted solution is described here. An external view and a vertical section of this SM are shown schematically in Figs.~\ref{fig:SM}.

After testing different designs satisfying the requirements~\cite{Anastasi2017},  a SM was built, characterized by a rigid mechanical structure, with a large thermal inertia and good electrical shielding which contains all detector components. The method employed to eliminate beam-pointing fluctuations relies on the use of an integrating sphere. An external view and a vertical section of this SM are shown schematically in Fig.~\ref{fig:SM}. 

\begin{figure}[htbp]
\centering
\includegraphics[width=0.42\textwidth]{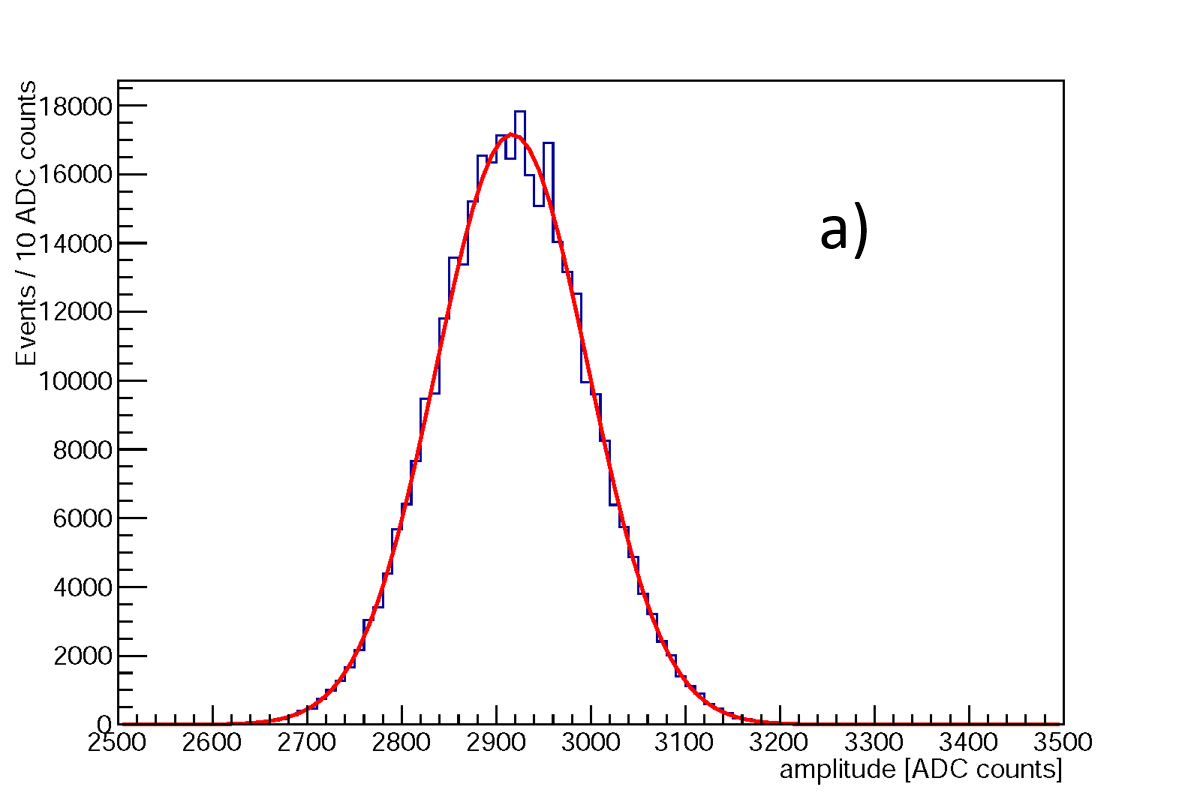}
\includegraphics[width=0.42\textwidth]{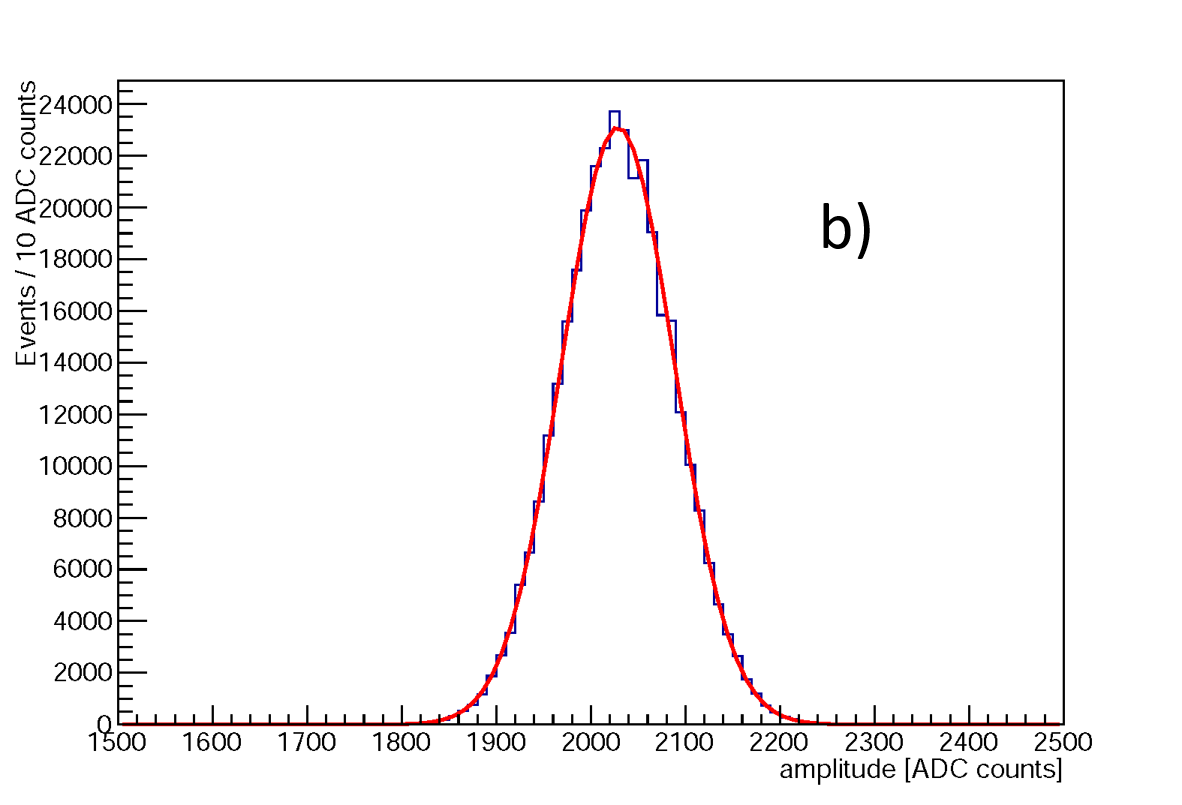}
\includegraphics[width=0.84\textwidth]{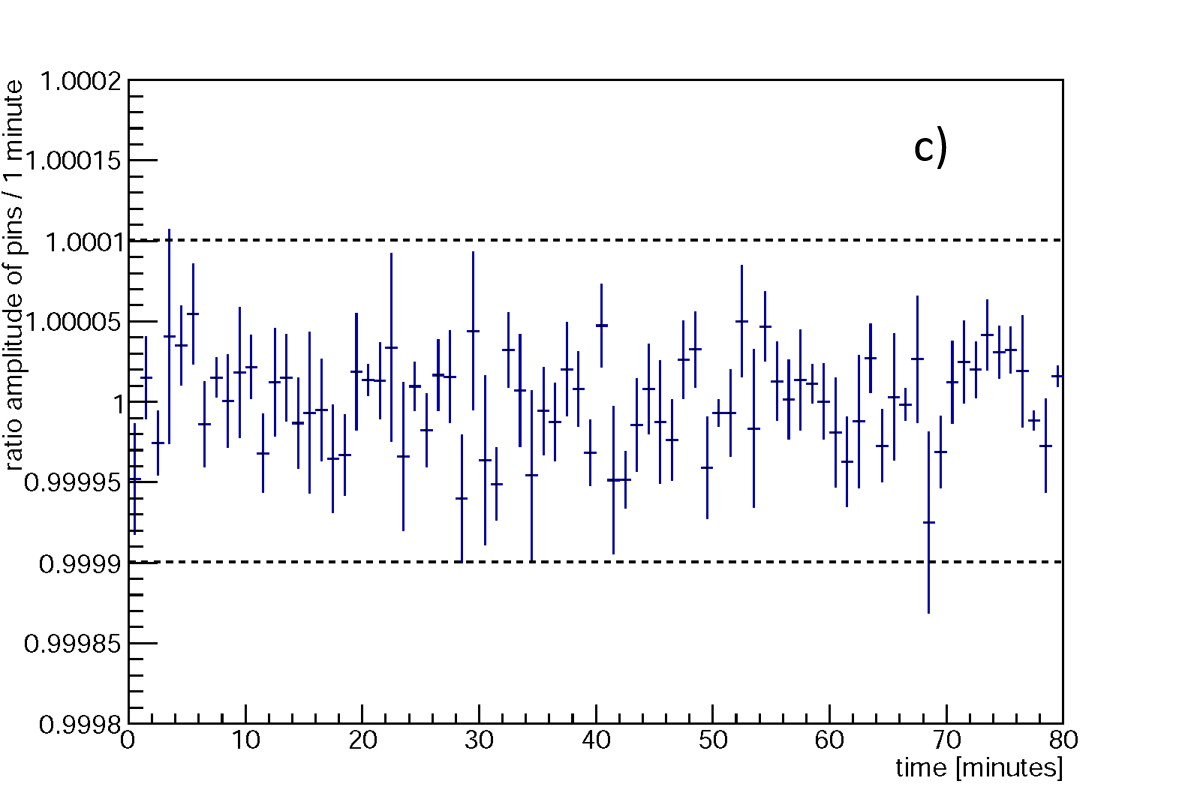}
\caption{Statistical distributions of laser signals from (a) the photomultiplier and (b) one of the PIN diodes, as measured at the input to the DAQ, i.e. at the output of the front-end electronics. (c) The variation of the ratio between the two PIN diode signals for SM1. Each point is an average of 3000 ratios and the two dashed lines indicate the  $10^{-4}$ stability limit over the 80-minutes acquisition period.}
\label{fig:laser_signals}
\end{figure}

The SM detectors and the integrating sphere are contained in a rigid aluminum case, to which a 30:70 (R:T) beam splitter is fixed by steel rods. The light intercepted by the splitter is focused onto the entrance port of the integrating sphere where it can be attenuated and diffused by appropriate optical elements. The light entering the sphere is distributed amongst three output ports where it is intercepted by 2 large-area  (1~cm$^2$) PIN diodes (PINs)~\cite{HamamatsuPD} and one Y11 wavelength shifter~\cite{wls}. The latter improves the matching of the laser wavelength to the photo-cathode of the SM photomultiplier (PMT) to which light is conveyed by a PMMA light-guide. This guide also accommodates a $^{241}$Am/NaI light pulser~\cite{Am_source,bicron} so that the PMT can detect both the signal originating from the laser source and a reference signal generated by the interaction of the $\alpha$ particles emitted by the $^{241}$Am source with the NaI in the light pulser.  This reference signal is used to correct for instabilities in the PMT response and, since both the PMT and the PINs receive the same laser signal, it can be used to correct for instabilities in the average PINs' response over sufficiently long periods to accumulate the required statistical precision. Optical and geometric parameters are adjusted so that the integral of the laser and reference signals are comparable and the wavelength shifter also helps to broaden the time-width of the laser signal, thereby reducing the difference in the dynamic range of the laser signal and that of the much broader reference signal and avoids saturating the electronics. 
%The shot-by-shot resolution of the SM signal (about 2.6\% as shown in Fig.~\ref{fig:signal_digitized}) is determined by the resolution of the PMT. 
The PINs see about 10$^6$ photo-electrons/laser pulse which corresponds to a statistical precision of about 10$^{-3}$ 
(0.1\%) per shot and the measured statistical resolution (about 0.3\%) is determined by the front-end electronics.  About 100 pulses (in nearly 0.02 seconds at an operational frequency of 5~kHz, used in dedicated calibration runs) are therefore sufficient for a statistical precision of better than 0.04\%. The stability of the PINs is illustrated in Fig.~\ref{fig:laser_signals}c. Given the PMT resolution (2.6\%), nearly 10$^{4}$ pulses (corresponding to 2 seconds) are required to attain a comparable statistical precision.

%\begin{figure}[htbp]
%\centering
%%\includegraphics[width=0.75\textwidth]{}
%\caption{…… qui’ ci andra’ la variazione dei segnali PiD in entrata al digitizzatore}
%\label{fig:signal_digitized}
%\end{figure}
%
The PINs are inherently stable, unitary gain, high speed devices which are insensitive to bias variation. Variations in the response are essentially due to temperature and correspond to 0.1\%/$^{\circ}$C~\cite{Anastasi2017,HamamatsuPD}. 
%However the temperature dependence illustrated in Fig.~\ref{fig:signal_digitized}a includes the effects of temperature variations on the front-end electronics. The temperatures reported in Fig.~\ref{fig:signal_digitized} are measured close to the SMs, inside the optical table enclosure. 

\begin{figure}[htbp]
\centering
\includegraphics[width=0.75\textwidth]{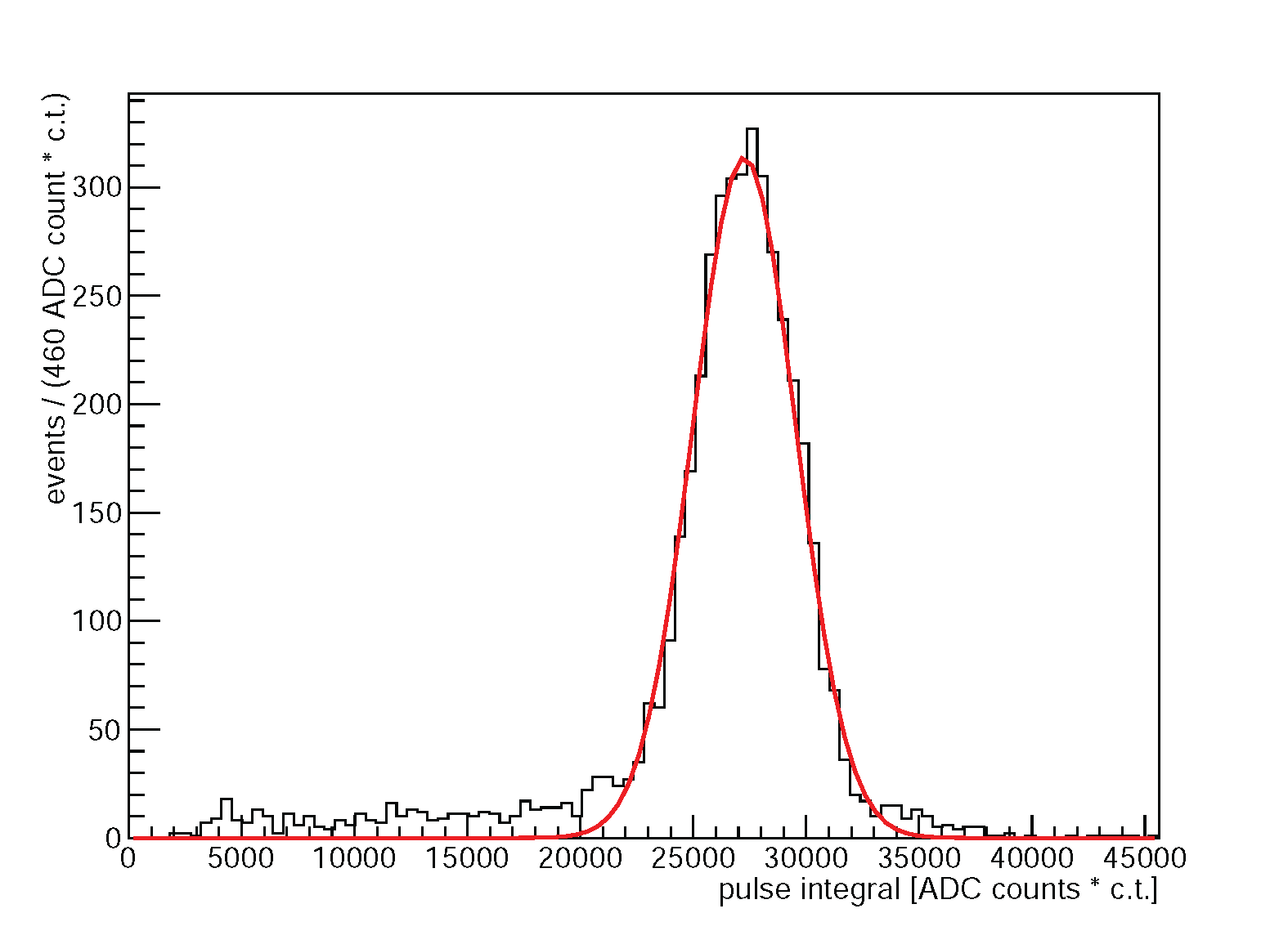}
\caption{The distribution of signals generated by the emission of 5~MeV $\alpha$ particles by the $^{241}$Am deposited on the surface of the NaI in the reference pulser. Notice the much larger distribution of pulse amplitudes with respect to the laser pulses of Fig.~\ref{fig:laser_signals}a.}
\label{fig:Americium}
\end{figure}

The PMT response depends both on the temperature and on possible variations of the PMT bias voltage. However, all variations of the PMT response that do not depend on the laser fluctuations can, in principle, be corrected by using the signals from the reference Americium pulser which the PMT views concurrently. Given the low activity (about 7~Bq) of the $^{241}$Am source~\cite{Am_source}, together with the distribution  ($\sigma$/mean $\approx 10$\%) (see Fig.~\ref{fig:Americium}) of the reference $\alpha$-particle signal generated by the pulser, long monitoring times may be required. As an example, about 30 minutes are required to monitor temperature variations of the PINs' response at a level corresponding to the PIN temperature sensitivity (0.1\%/$^{\circ}$C) which, given the rate of temperature related variations of the PINs signals, is sufficient to parametrize these variations.

Data from the SM detectors are recorded by both the main DAQ system, through the waveform digitizers used by the calorimeters, and by an independent, local DAQ as described in Section~\ref{ssec:processing}.

\subsection{The Local Monitor}
\label{ssec:local}
%(Giovanni Cantatore, Trieste)
%\subsubsection{The supermonitor?}
%\label{sssec:supermonitor}

The Local Monitor is a component of the calibration system designed mainly to monitor the
stability of the light distribution system. It consists of $2 \times 24$ photo-multiplier tubes (Hamamatsu R1924A,~\cite{HamamatsuR1924A}) receiving two optical signals back from each one of the 24 calorimeters\footnote{In the first run of the experiment only 24 PMTs were installed, subsequently the system was doubled for redundancy.}.  The PMTs are mounted in three ventilated boxes together
with shaping and adapting electronics for the final readout by waveform digitizers. The data are then sent to the experiment DAQ and stored.

\begin{figure}[htbp]
\centering
\includegraphics[width=0.48\textwidth]{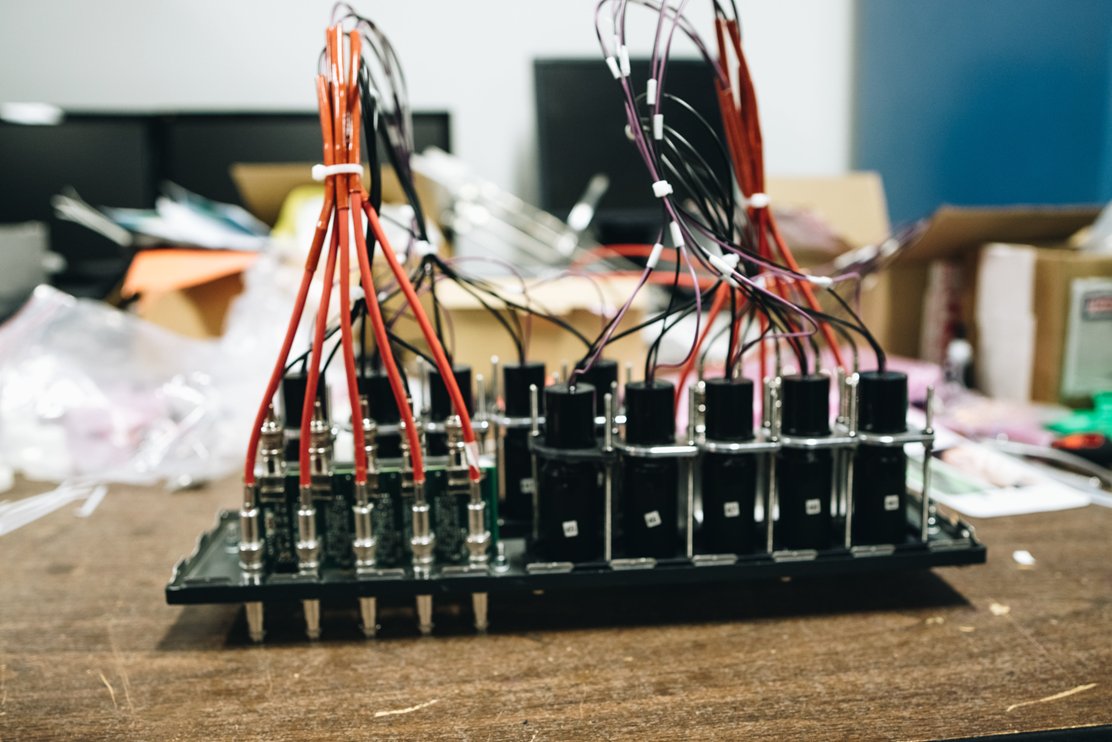}
\includegraphics[width=0.48\textwidth]{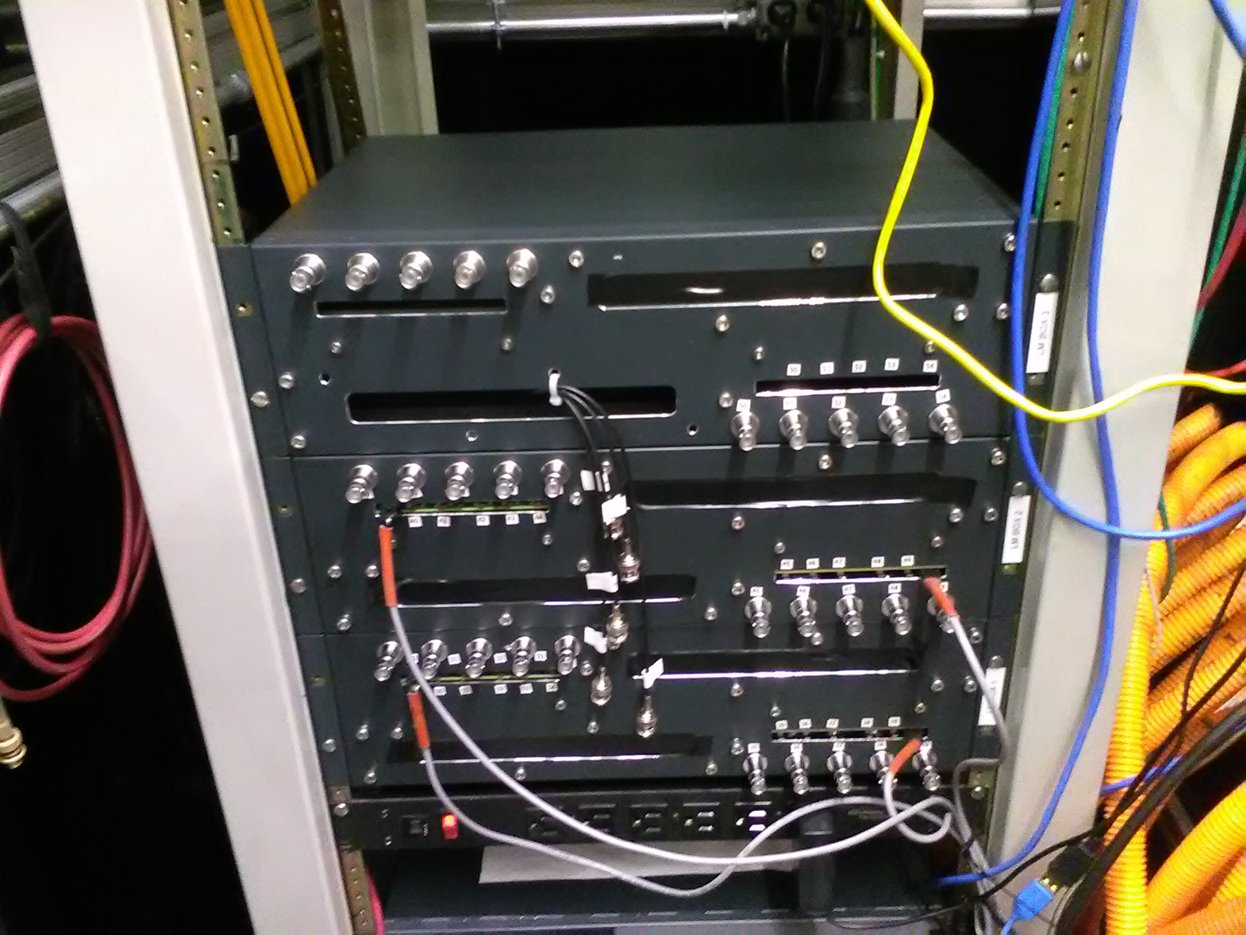}
\caption{Details of the LM. Left picture: PMTs, electronics and connectors are fixed on the front panel. Right picture: LM boxes in the rack prepared for connecting the fibers, the fiber receptacles are covered
with black tape and, after connection, by a light-tight plastic conduit.}
\label{fig:LM_photos}
\end{figure}

The LM boxes are
placed in the thermally controlled Laser Hut, in racks close to the optical table. Close to the boxes
there is also a CAEN 127 HV supply that provides the necessary voltage for the PMT operation. The LM boxes and rack are shown in Figs.~\ref{fig:LM_photos}.

As described in Sec.~\ref{sec:distribution}, calibration laser pulses are directed from the optical table to each of the 24 calorimeters by launching silica fibers, each coupled to a diffusing box. In principle all elements of the distribution chain, the optical elements to split the beam in 4 and inject the fibers, the launching fibers, and the diffusing box, are designed specifically to ensure the necessary stability of the laser light delivered to the calorimeter. Nevertheless either small changes, mainly due to temperature variations, or abrupt ones due to catastrophic events like fiber damage may occur during the long run-time period. The LM is thus intended to give a prompt diagnostic of the status of the distribution chain and provide a correction to the SiPM response due to variations in the light distribution chain.  This is achieved by taking a small fraction of light from the diffuser with two long fibers (one plastic and one silica)  and sending them back into the Laser Hut to two PMTs. The redundancy in the LM is needed to study and compensate for fluctuations due to temperature of the transmission coefficient of the LM optical fibers. The use of two different types of fibers allows the monitoring of temperature and solarizing effects on the PMMA
fibers.

\begin{figure}[htbp]
\centering
\includegraphics[width=0.85\textwidth]{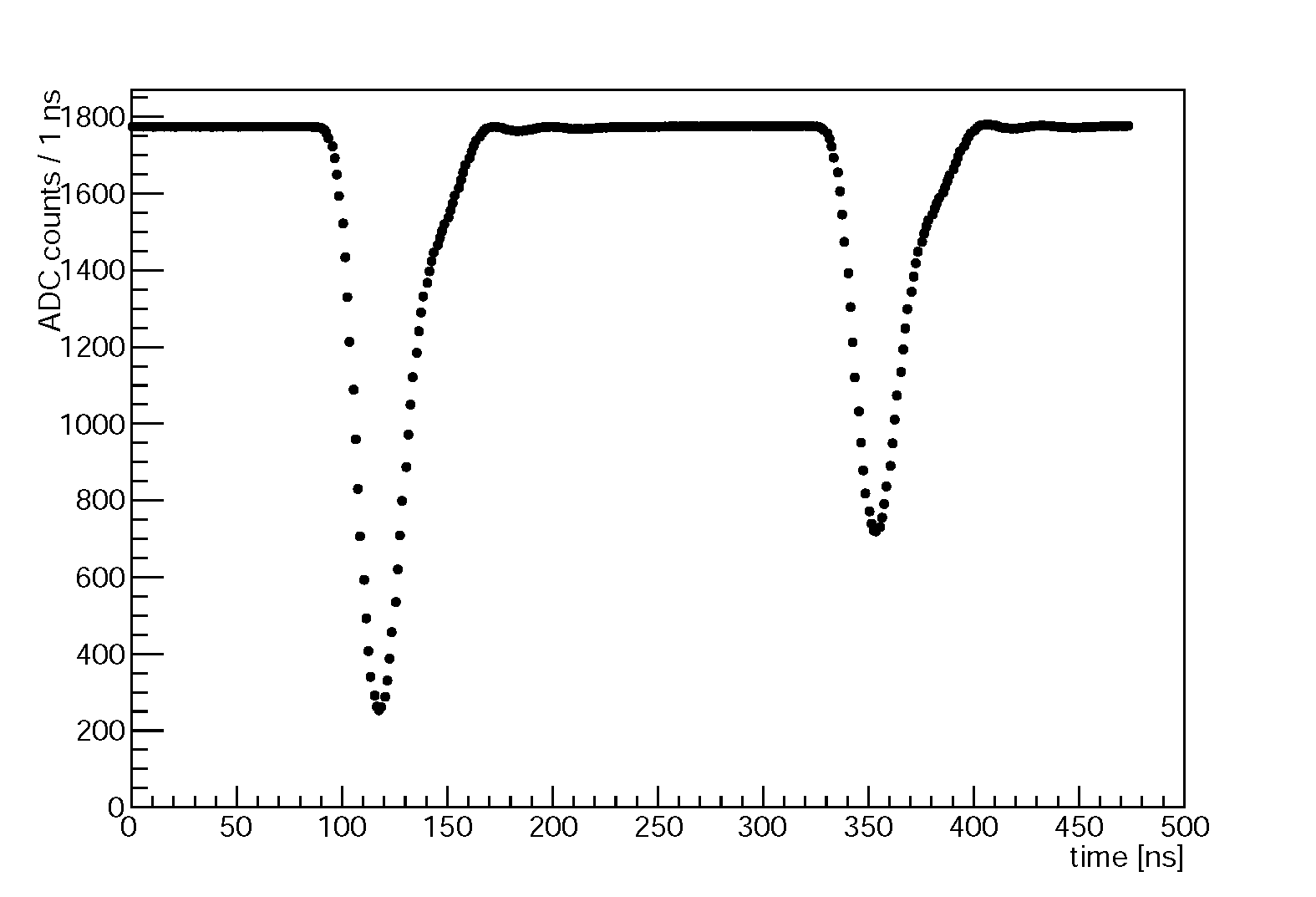}
\caption{Typical trace of a Local Monitor PMT signal. Notice that the height of the delayed LM peak is  only incidentally smaller than the reference SM one, as both peak intensities can be independently adjusted.}
\label{fig:K3}
\end{figure}

In order to be independent of
possible fluctuations of the PMT gain in the LM, a small quantity of light is also taken directly from the integrating sphere of the Source Monitor and  guided by an optical fiber to each PMT of
the LM to provide a reference signal (each SM feeds the 4 LMs corresponding to the 4 calorimeters that are illuminated by the same laser head). In this way each PMT sees two pulses (see Fig.~\ref{fig:K3}) separated by roughly 240~ns,
%a delay proportional to the sum of the lengths of the ingoing and outgoing optical fibers; we neglect the length of the fiber going
%from SM to LM since it is an order of magnitude shorter. The delay is about  long,
corresponding to the 50 meters of fiber going back and forth from the Laser Hut to the calorimeter position.  

The PMTs with their magnetic shielding are fixed on the box front panel, which is made of black Delrin,
and the fibers (both short and long ones) are glued in the Delrin plugs used for their fixation. The fibers are 1~cm distant from the PMT photo-cathode in order to allow the placement of
an interference band-pass filter centered at 405 nm in between, and let the beam expand to about 1 inch diameter before hitting the PMT. In this way only the light at the laser's
wavelength contributes to the signal and the beam covers an important fraction of the photo-cathode
thus averaging its local properties. 
%The Delrin front panel of the PMT box with the bandpass filters mounted is shown in Fig.~\ref{fig:K4}
%
%\begin{figure}[htbp]
%\centering
%\includegraphics[width=0.75\textwidth]{Karuza_4.png}
%\caption{Picture showing one of the front panels of a Local Monitor box. Two blocks of 5 PMTs are hosted inside each box, each PMT entrance window is protected with a Thorlabs FB405-10 bandpass filter to reduce room scattered light noise.}
%\label{fig:K4}
%\end{figure}
%

In the LM, 24 out of 48 PMT signals are split and recorded also by a local data acquisition system.  
%The redundancy in the LM is needed to study and compensate for any fluctuations due to temperature of the transmission coefficient of the LM optical fibers. This is achieved also by using two types of fibers for the light coming back from the ring: quartz and PMMA. The system is redundant and allows the monitoring of temperature and solarizing effect of the PMMA fibers.

\section{The control electronics and the readout system}
\label{sec:electronics}
%control electronics (Stefano Mastroianni, Napoli)

\begin{figure}[htbp]
\centering
\includegraphics[width=0.95\textwidth]{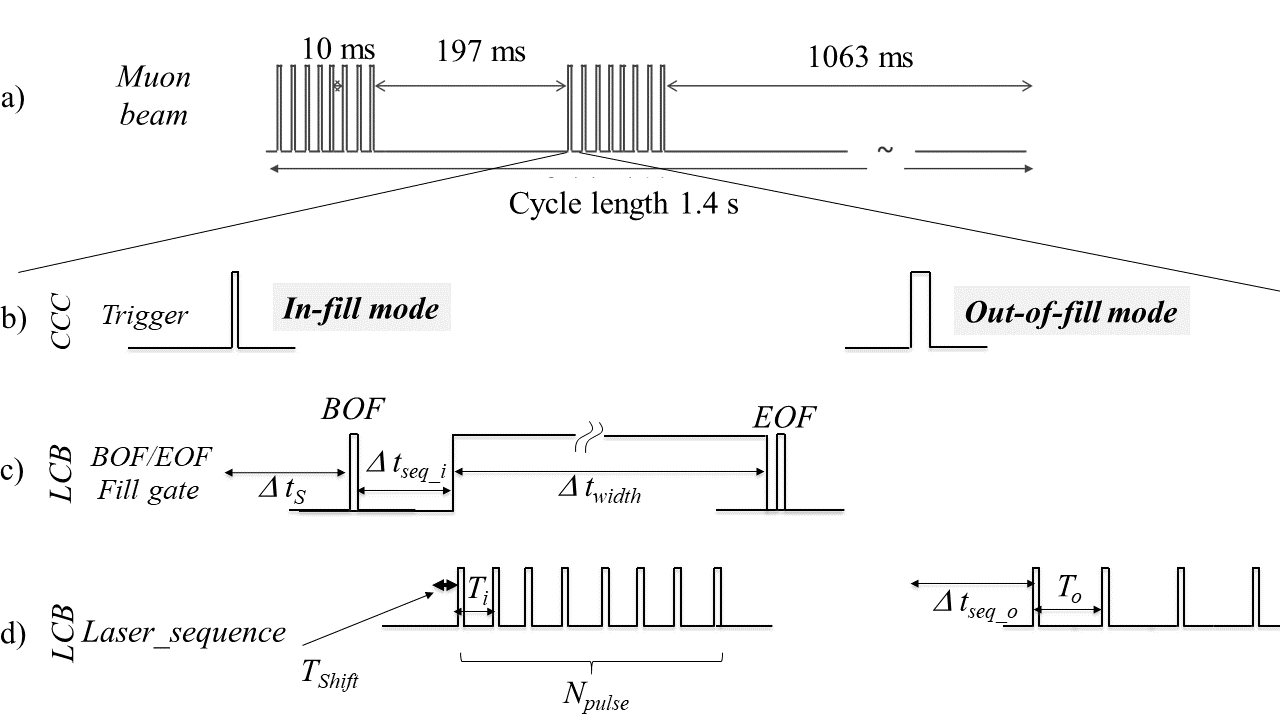}
\caption{Time sequence of the different trigger pulses generated for each muon injection. In (a) the time sequence of the complete beam-machine cycle ($1.4\,$s period) is shown, with each square pulse representing a muon fill. For each muon fill, the CCC generates one start (In-Fill Mode) and one stop (Out-of-Fill Mode) triggers (b) that are used by the LCB to produce Begin-Of-Fill and End-Of-Fill triggers (c) and, eventually, the triggers for the generation of the laser pulses, IFP and OoFP respectively (d). The BoF pulse is generated by the LCB a time $\Delta t_{\rm S}$ after the CCC In-Fill Mode trigger and thus precedes by $\Delta t_{\rm seq-i}$ the beginning of the muon fill. The EoF pulse is generated by the LCB after the end of the muon fill, lasting $\Delta t_{\rm width}$ (700$\mu$s). $T_{\rm i}$ and $T{\rm o}$ are the pulse separations of the IFPs and OoFPs respectively, $T_{\rm Shift}$ is the delay of the IFP series with respect to the beginning of the muon fill and $\Delta t_{\rm seq-o}$ sets the delay of the OoFP series with respect to the CCC Out-of-Fill Mode trigger.}
\label{fig:Pulse_generation}
\end{figure}

Specific electronics modules have been designed to manage the complete control of the laser system and photo-detector data readout, as well as to provide bias voltage and control signals. 
A key element of the laser calibration system is the Laser Control Board (LCB) that manages the interface between the beam cycle and the calibration system, takes care of the generation of the laser pulses and distributes the time reference signals to the monitoring electronics (see the complete trigger sequence in Fig.~\ref{fig:Pulse_generation}). 
Although all SM and LM signals are digitized by the WaveForm Digitizers (WFDs) and  stored in the general DAQ of the Muon $g-2$ experiment, a local calibration system DAQ with  a modular structure, has been developed. It is based on an event-driven data collection and uses a custom bus protocol and a controller board for the readout from Monitoring Boards (MBs). Each crate contains up to 12 such MBs managing 36~photo-detector channels and the controller.

\subsection{The Laser Control Board}
\label{ssec:control}

The laser is operated in four main different modes, as described in Sec.~\ref{sec:operation}. The first is set during physics runs and allows the correction of  systematic effects due to gain variation of the SiPMs, produced by the very high muon decay rate during the $700\,\mu$s muon fill. The second mode corresponds to the double-pulse mode, used to study SiPM behavior to two consecutive pulses. The third one is used to equalize the gain of the different SiPMs inside a calorimeter and, finally, the fourth allows the simulation of physics events produced by the muon decay. It is enabled for the calibration runs, without beam, in order to test the detector, front-end electronics and DAQ. Moreover, the laser is used for time alignment of the SiPM in the calorimeters and between calorimeters. 

In these operation modes the LCB provides~\cite{Anastasi2018}:

\begin{itemize}
  %	\item pulses train generation, OoFP and IFP respectively, at programmable frequency  in in-fill and out-of-fill gaps of  the beam cycle during  physics runs;
  	\item In-Fill and  Out-of-Fill Pulse generation at a programmable frequency during  physics runs;
	\item single pulses or pulse train generation at a programmable frequency in special {\it double-pulse} runs (no beam present);
	\item single pulses at a  programmable frequency for calibration purposes (no beam present);
	\item physics event simulation, or operation in Flight Simulator Mode (FSM), by triggering the laser according to the exponentially decreasing time function, $\exp (-t/\tau)$ , as expected in the experiment due to muon decay; many other distribution functions can be simulated to reproduce a specific effect in detector and DAQ system.
	\item time reference signal for synchronization and initialization of the detector and front-end electronics and DAQ. 
\end{itemize}

The LCB manages the interface between the calibration system and the experiment's synchronous control system, the Clock and Control Center (CCC). The CCC provides the triggers to the LCB timed appropriate to delivery of the muon beam. The LCB decodes the trigger mode and generates the suitable laser pulse sequence.

During the physics runs, the generation of laser pulses both for in-fill and out-of-fill time windows is mainly based on a programmed number of electronic trigger pulses issued during an enable gate.  The programmable pulse rate  spans from hundreds of Hz to MHz. This mode is completely implemented in hardware and it needs only to be configured.
In simulation mode (FSM, see Sec.~\ref{ssec:flight}) the LCB is able to repeatedly provide a time sequence with a
mean number of pulses corresponding to an exponential positron-decay-time
distribution by means of an exponential function $\exp (-t/\tau)$  with a decay time of 64.4~$\mu$s. 

\begin{figure}[htbp]
\centering
\includegraphics[width=0.75\textwidth]{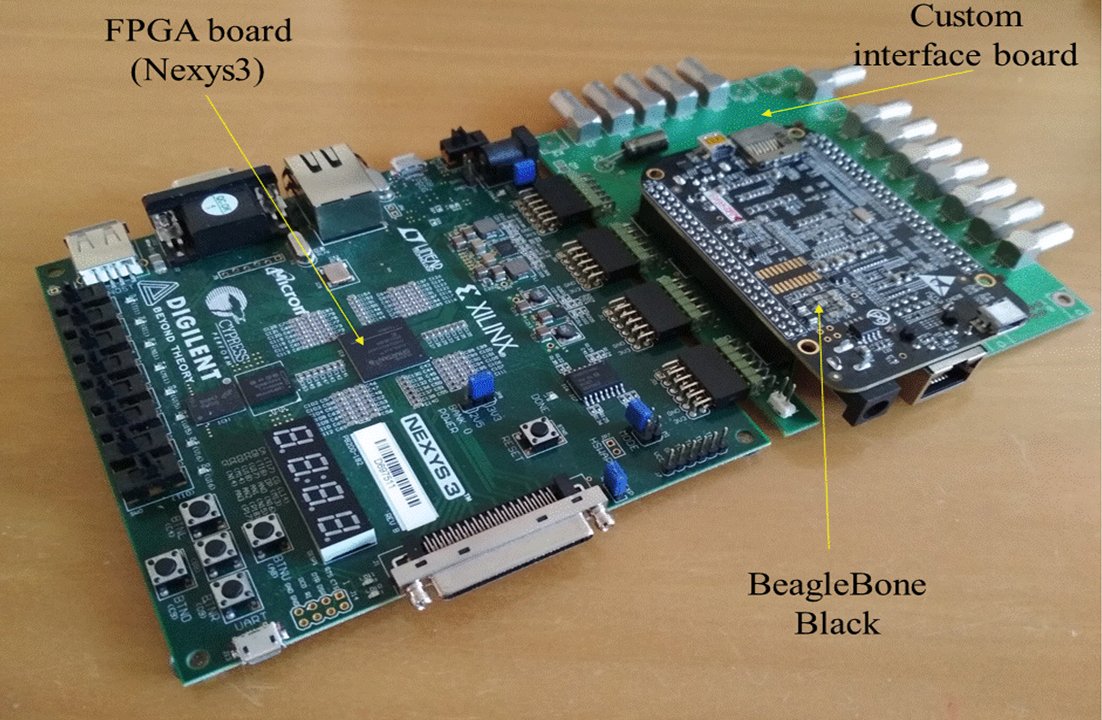}
\caption{Picture of the LCB, which is implemented by a hybrid platform hosting a Spartan6 FPGA board and an embedded CPU}
\label{fig:Na_board}
\end{figure}

\begin{figure}[htbp]
\centering
\includegraphics[width=0.95\textwidth]{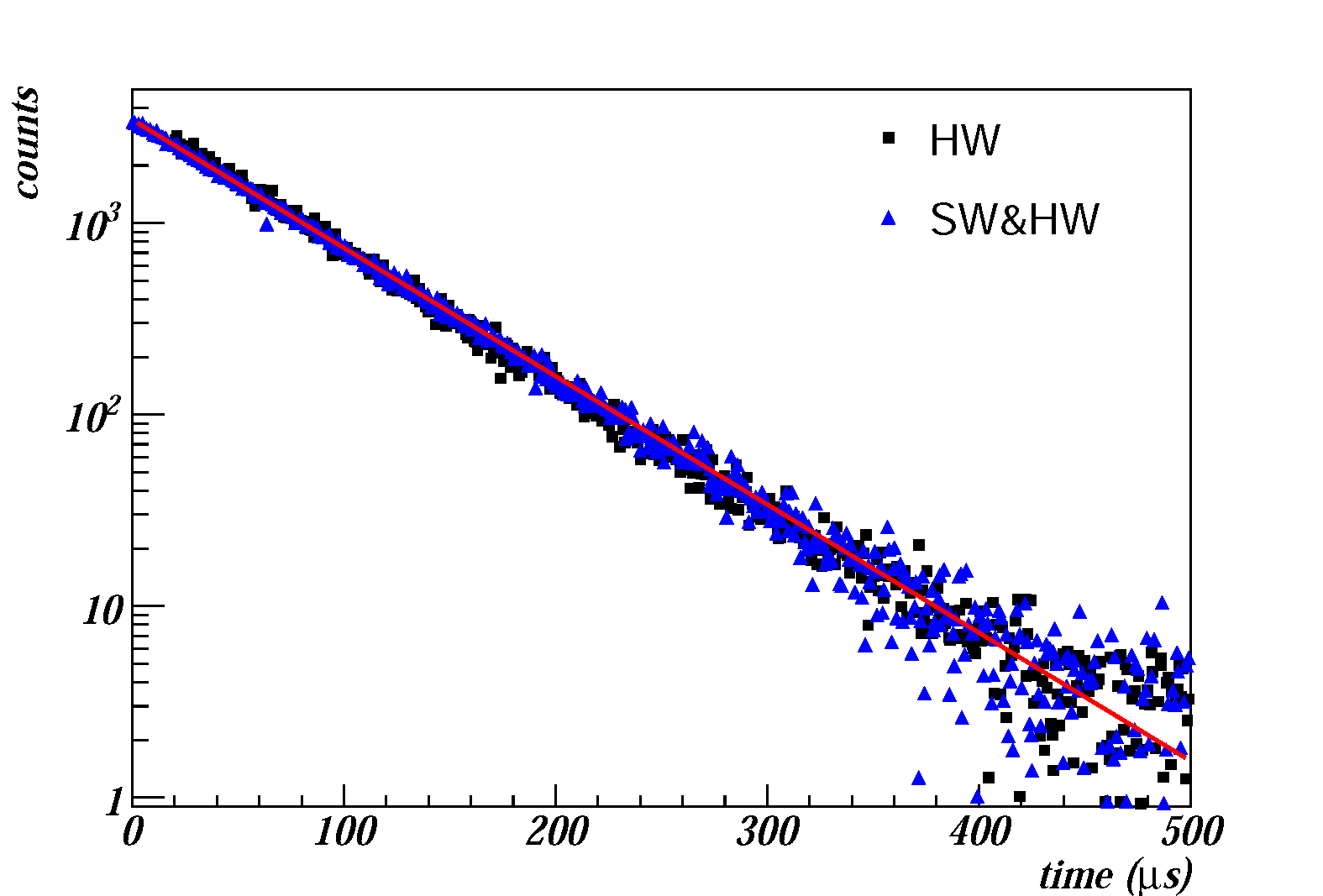}
\caption{Time distribution of laser pulses generated in the ``flight simulation'' mode, according to the two implemented methods. Both distributions nicely follow and exponential decay law.}
\label{fig:Pulse_distribution}
\end{figure}

The system consists of a hybrid platform hosting a FPGA board and an ARM-based processor (Fig.~\ref{fig:Na_board}). The use of an embedded processor has the advantage of hosting a high level operating system. Two different implementations of FSM have been built. The first implementation (HW) is completely based on the hardware, which can be controlled by modifying the firmware inside the FPGA device.  The second solution (SW and HW) contains a pulse generator controlled by an embedded processor. Fig.~\ref{fig:Pulse_distribution} shows a good agreement between the distributions of pulses obtained using the two implementations. This version has been used to do several tests on the SiPMs~\cite{Kaspar2017} and was recently used in different test beams, in particular at Laboratori Nazionali di Frascati~\cite{Anastasi2017} and at SLAC National Accelerator Laboratory~\cite{Khaw2019}. The capability of simulating a specific time distribution function is an interesting benefit useful for detector and DAQ measurements. The entire system is managed and controlled remotely including updating of the firmware inside the FPGA device. 

%The system has been realized~\cite{Anastasi2018} by using a hybrid platform with FPGA board and ARM-based processor. The use of an embedded processor has the advantage to host an high level operative system.  Fig.~\ref{fig:Na_board} shows the implementation based on a Xilinx Nexys board and a BeagleBone ARM processor). Several tests have been done to validate all the pulse generation modes.  The LCB contains two different implementations of the flight simulator. The first (HW) is fully realized in hardware (FPGA) while the second (SW and HW) consists in a hardware generator of pulse sequences whose patterns are provided by software modules (CPU). Fig.~\ref{fig:Pulse_distribution} shows a good agreement between the distributions of pulses obtained using the two implementations.
%This version has been used to do several tests on the SiPMs~\cite{Kaspar2017} and was recently used in different test beams in particular at LNF~\cite{Anastasi2017} and at SLAC. The possibility to simulate a specific time distribution function is an interesting benefit useful for detector and DAQ measurements. All the system is fully managed and controlled remotely including the firmware updating inside the FPGA device. 

\subsection{Signal processing and data readout}
\label{ssec:processing}
%end of control electronics

The Monitoring Board~\cite{Anastasi2019} has been specifically designed to manage the complete signal processing, data readout and configuration/control for one Source Monitor element made of two PIN photo-diodes and one PMT. The MB has three independent sections to manage the three signals.  The first stage of the chain is a preamplifier circuit. The output is sent to a pulse shaper that transforms the signal with a long tail (about 20~$\mu$s, determined by the RC time constant of the preamplifier) to a semi-Gaussian shape around the peaking time. A baseline restorer circuit is used to remove the baseline shift that would cause an uncertainty in the peak determination. Shaping and baseline restoration circuits feed an analog to digital converter (ADC).  A peak detector is used to track and hold the peak value long enough for the digitization process. In addition, the MB  provides a filtered duplicate of the three signals from the three photo-detectors to allow an independent, redundant digitization by the $\mu$TCA-based WFD boards of the main DAQ of the experiment. The preamplifier circuit has been assembled on a daughter board placed near the sensor to reduce the electronic noise. The MB allows the use of a single hardware platform to manage different photo-detectors. In fact, the module is based on FPGAs and can be customized by means of the configuration files loaded for the Local Monitor photo-detectors.

%MB has been specifically designed to manage the complete signal processing, data readout and configuration/control for one Source Monitor element made of 2 PIN photodiodes and one PMT. The board has three independent sections to manage the three signals.  A preamplifier circuit is the first stage in the processing chain. Then, a pulse shaper transforms the output from the preamplifier with a long tail (about 20~$\mu$s) to a semi-Gaussian shape around the peaking time. A baseline restorer is used to avoid the baseline shift causing an uncertainty in the peak determination. Shaping and baseline restoration circuits feed an analog to digital converter (ADC) for digitization process.  A peak detector is needed to track and hold the peak value long enough to allow the quantization process. In addition, the MB module provides the three filtered signals of photo-detectors to the WFD boards to allow an independent digitization based on $\mu$TCA crate. To improve the flexibility of electronics chain, the preamplifier circuit has been implemented on a daughter board near the sensor and connected with a data/control flat cable to the MB. This solution allows the use of a unique hardware platform to manage different photo-detectors. In fact,  the module, based on FPGAs, can be customized by means of the loaded configuration files for the Local Monitor photo-detectors.

The local data acquisition system is built on a trigger-driven scheme where all MBs and Controller~\cite{Mastroianni2018} share the same trigger signal distributed by the LCB.  On the arrival of a trigger, each MB performs the data framing by collecting all the sub-frames (containing baseline measurement, peak value and time information of each signal) pertaining to the same trigger, adds slow control information (temperature and bias measurements) and other control words and pushes the reconstructed frame into a FIFO implemented in the FPGA. The frame is then transferred to the Controller which in turn performs the event building at crate level. All the sub-frames from MBs belonging to the same trigger number, after the integrity data checks, are stored into a FIFO accessible by an embedded processor for the final readout.

%The Controller takes care of the data collection from the 6 Monitoring Boards. 
The Controller can take care of the data collection from a maximum number of 12 Monitoring Boards. In the present configuration we use only one Controller connected to 6 MBs.
Each MB is connected to a Controller by means of two unidirectional serial links. There are also some control signals (trigger signal, synchronization/reset signals, busy) that are broadcast by the Controller to slave boards (in this case the 6 MBs) to implement the communication protocol. 

\begin{figure}[htbp]
\centering
\includegraphics[width=0.75\textwidth]{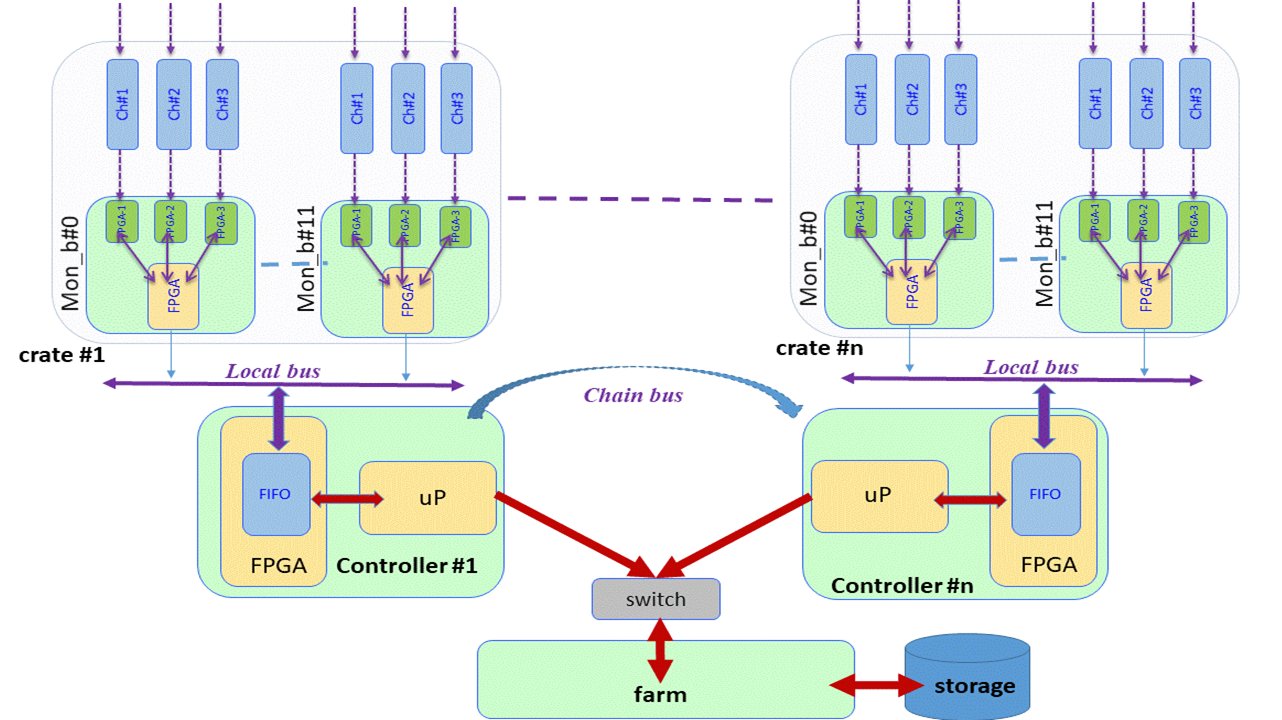}
\caption{Schematics of the locally implemented data acquisition system, based on multiple crates.}
\label{fig:DAQ_chain}
\end{figure}

The data transfer between slave boards and Controller is based on an RS-232-inspired serial-link scheme. The input section of the Controller has 12 identical slices to manage the readout from slaves; in particular it reconstructs each slave frame, checks its integrity, and writes it into a buffer FIFO. The use of a buffer memory improves the decoupling between the input and output sections.

%When all the data frames from the slave boards pertaining to the same trigger number are written in the buffer FIFO, the event building block starts the parsing of sub-frames and stores them into the building FIFO until the last board is reached. 
When all the data frames from the slave boards pertaining to the same trigger number
are written in the {\it buffer} FIFO of the Controller, the event building block starts the
parsing of frames and stores them into the {\it building} FIFO of the Controller until the
last board is reached.
An embedded processor hosted on the board manages the final readout by using a USB interface and transfers the data to the online farm for further processing.

%The data acquisition system is based on an trigger-driven algorithm where all MBs and Controller share the same trigger signal coming from the LCB. When a trigger arrives, each MB board performs the data assembling by collecting all the pulse sub-frames from the three channels of the sub-cycle, builds a header with control words and slow control information (temperature and bias measurements) and labels the acquired data with the respective event number. Such a reconstructed frame is pushed in a FIFO implemented in the fourth FPGA and then transferred to the Controller which in turn performs the event building at crate level. It processes all the sub-frames from MB slaves pertaining to the same trigger number, checks the data integrity, adds control and monitoring words and stores the frame in a FIFO accessible by an embedded processor for the final readout.

%The Controller manages data collection from a maximum number of 12~slave boards. Each board is connected to a Controller by means of two unidirectional serial links (to send and receive). There are also some control signals that are broadcast by the Controller to slave boards to implement the communication protocol (i.e. trigger signal, synchronization/reset signals, busy). Each slave, in case of an error condition occurred during the data taking, can assert a stop signal on a wired-OR line and the Controller starts a readout cycle to register the status of the boards.

Several Controller boards can be chained together in case of multiple crates, as shown in Fig.~\ref{fig:DAQ_chain}. The event building at crate level is fully implemented in hardware, while the event building at chain level must be implemented at farm level. At present, the laser calibration system of the Muon $g-2$ experiment consists of 2 crates, one for Source Monitor and another one for the Local Monitor.

\section{Laser operation modes and calibration procedures}
\label{sec:operation}
%description of the different calibration procedures (Graziano Venanzoni, Pisa)

The laser calibration system has been used in several ways during the commissioning and the running periods. The functioning modes can be set remotely from the control room at the beginning of each run (by {\it run} we intend here any uninterrupted period of data taking with a given set of experimental conditions in the whole $g-2$ system; it can last from a few minutes to few hours and data are recorded in an unambiguous way in the main data stream by the DAQ). Some of these functioning modes have become routine and are described in the following: the {\it standard}, the {\it double-pulse}, the {\it gain calibration}, and the {\it flight simulation} operation modes.

\subsection{Standard operation mode}
\label{ssec:standard}

The {\it standard operation mode} is the working procedure that provides gain corrections to calorimeter signals  and  time-reference for synchronization of different detectors, front-end electronics and DAQ. The time sequence of the laser pulses, triggered by the Laser Control Board, has been shown in Fig.~\ref{fig:Pulse_generation}. As shown, muon fills from the FNAL line arrive in bunches of eight, at 100~Hz frequency, each fill lasting about 700~$\mu$s. In the standard mode, before the arrival of the muon fill, a first {\it Sync/BoF} laser pulse is fired for synchronization, then a number of IFP are fired within the time of the fill, separated by 200~$\mu$s and delayed with respect to the beginning of the fill by a variable time lag $T_{\rm Shift}$. A time $\Delta t_{\rm seq-o}$ after the end of the fill, four OoFP  are fired, separated by a time lag $T_{\rm o}$.

\begin{table}[htbp]
	\centering
		\begin{tabular}{|l|l|l|l|}
		\hline
		  Error				& E821		&		E989								& 	Goal		\\
			Category		& (ppb)		&		improvement plans		&		(ppb)		\\	
		\hline
			Gain changes	&		120		& 	{\bf Improved Gain Monitoring}								& 		  \\
%			Gain changes	&		120		& 	{\textcolor{red}{Better laser calibration}}		& 		  \\
										&					&		low-energy threshold												&		20	\\
			Pileup				&		80		&		low-energy samples recorded									&				\\
										&					&		calorimeter segmentation										&		40	\\
%			Lost muons		&		90		&		better collimation in ring									&		20	\\	
%			CBO						&		70		& 	higher {\it n} value												&				\\
%										&					&		better match of beamline to ring						&		<30	\\
%			E and pitch		&		50		&		improved tracker														&				\\
%										&					&		precise storage ring simulations						&		30	\\
		\hline
%			Total					&		180		&		quadrature sum															&		70	\\
%		\hline			
		\end{tabular}
\caption{Systematic uncertainties for gain changes and pileup for the BNL E821 experiment $\omega_{a}$ analysis and  for the FNAL E989~\cite{Grange2015}. The goal for the {\it total} uncertainty in the latter experiment is 70~ppb.}
\label{tab:errors}
\end{table}

Fluctuations in the gain stability of the calorimeter system and {\em pileup}
effects, i.e. random overlap of different positrons in the same
calorimeter, are the main sources of systematic error for the
measurement of the precession frequency, see Table~\ref{tab:errors}. The goal of the  standard operation mode is to
send a regular pattern of laser pulses which are then used offline to
calibrate the system. In particular, three sets of pulses are issued, which are represented in Fig.~\ref{fig:Pulse_generation}: 

\begin{itemize}
	\item SYNC: a Begin-of-Fill (BoF, also known as SYNC) and an End-of-Fill 
	(EoF) pulse are sent to all 1296 calorimeter channels a
  few tens of microseconds before and after muon injection in the ring, respectively. These
  signals are used to synchronize the response of the crystals in such a way that the final accuracy of the time reconstruction is at the
  $\simeq 30 \, {\rm ps}$ level.
	\item IN FILL:  During a prescaled subset of  muon
	fills, the laser system fires a fixed number of
  pulses.  The pulses are shifted in time, for each subsequent fill, in order 
  to sample all times from the injection time up to several hundred
  microseconds later. This procedure will be described in detail in
  Sec.~\ref{sssec:in-fill}.
	\item OUT OF FILL: muon injections are interleaved by a time gap of
  $\simeq 10$~ms which
  allows the laser system to send a set of pulses, normally 4, when no muons are present. 
  These pulses are used 
  as a long term stability check of the calorimeters, and in
  particular of the SiPMs' gain. 
  As described in Sec.~\ref{ssec:performances}, this so-called {\em Out-of-Fill Gain} (OoFG) correction 
  equalizes the  SiPM response as a function of time, as these devices
  are very sensitive to environmental fluctuations, in particular to temperature.

\end{itemize}

%\begin{\itemize}
%\item SYNC: a Beginning Of Fill (BOF, also known as SYNC) and an End
  %Of Fill (EOF) pulse are sent to all 1296 calorimeter channels 
  %few microsends before and after muon injection in the ring. These
  %signals are used to synchronize the response of the crystals at the
  %$\simeq 100 \, ps$ level.
%%\item IN FILL:  During a prescaled subset of the muon injections,
%%referred to as "fills", the laser system fires a fixed number of
%%pulses.  The pulses are shifted in time each subsequent ``laser fill''
%%to sample fill times from injection time to several hundred
%%microseconds later. This procedure will be described in detail in
%%section 5.5.2 .
%%\item OUT OF FILL: muon injections are interleaved by a time gap of
  %%$\simeq 10$ ms which
  %%allows the laser system to send a set of pulses (normally 4). 
  %%These pulses  are
  %%used 
%%as a ``long term'' stability check of the calorimeter, and in
%%particular of the Silicon Photomultipliers (SiPM). 
%%As described
  %%in section 5.5.3, this so-called {\em Out Of Fill correction} 
  %%equalizes the  SiPM response as a function of time, as these devices
  %%are very sensitive to
  %%environmental fluctuations, in particular to temperature. 
%\end{itemize}

The exact time sequence of these pulses is controlled by the Laser
Control Board and its parameters are set in the Online DataBase 
interface and stored for each run. 

\subsection{Double-Pulse operation mode} 
\label{ssec:double}

In the {\em Double-Pulse mode} (DP) two consecutive laser pulses
are sent to all crystals with a 
delay that can vary from 1~ns up to several hundreds of $\mu$s.

The goal of this mode is to test the 
calorimeter response to two or more
consecutive particles. 
From previous studies~\cite{Kaspar2017}, it is known that the
SiPM gain is reduced when two particles enter a crystal within a short time interval. 
This behavior is shown in Figs.~11 and~12 of Ref.~\cite{Kaspar2017} showing that the 
calorimeter response, or gain function, is not
flat as a function of time but it has two distinct time structures:
one at short time separations ($\sim 20$~ns),
due to SiPM response, and one at longer times ($\sim 20\, \mu$s), due to the recovery time of the power supply. 
The short time response curve has been measured in the laboratory
with LEDs, while the long time response
has been obtained with a SPICE simulation.
 Details can be found in the cited article.
The DP mode provides the 
 possibility of checking periodically (each three days of data taking, typically) the gain function for each of
 the 1296 crystals
during data taking which allows for correction of these effects and 
for keeping the systematic error under control.

%The gain function can be measured using the laser system, 
%which has the possibility of sending 
%two consecutive pulses, or bursts of pulses, to the same
%crystal with a programmable delay.
%The two pulses are generated by two different lasers illuminating the same crystal. In order to do this, a set of movable mirrors has been installed, as described in the next section.

There are several different reasons why it is better to send pulses from two
different lasers, rather than fire the same laser repeatedly:
\begin{itemize}
\item two lasers allow for choosing different light intensities for each pulse of the pair;
\item the laser maximum repetition rate of 40~MHz does not allow the
  testing of nanosecond time scale;
\item in case of two consecutive pulses, 
the laser output light for the second one can be
  systematically different from the first one, 
while light fluctuations for different lasers are uncorrelated;
\item as the laser light output can fluctuate up to the percent level  
from pulse to pulse and this fluctuation is 
  monitored by the SM (response time  tens of microseconds), it
   cannot be corrected when pulses are too close in time.
\end{itemize}

For all these reasons, the laser optics has been modified to include the
possibility of sending 2 different laser pulses to the same
calorimeter, as described in section~\ref{ssec:doublepulse}.

\begin{figure}[htbp]
\centering
\includegraphics[width=0.85\textwidth]{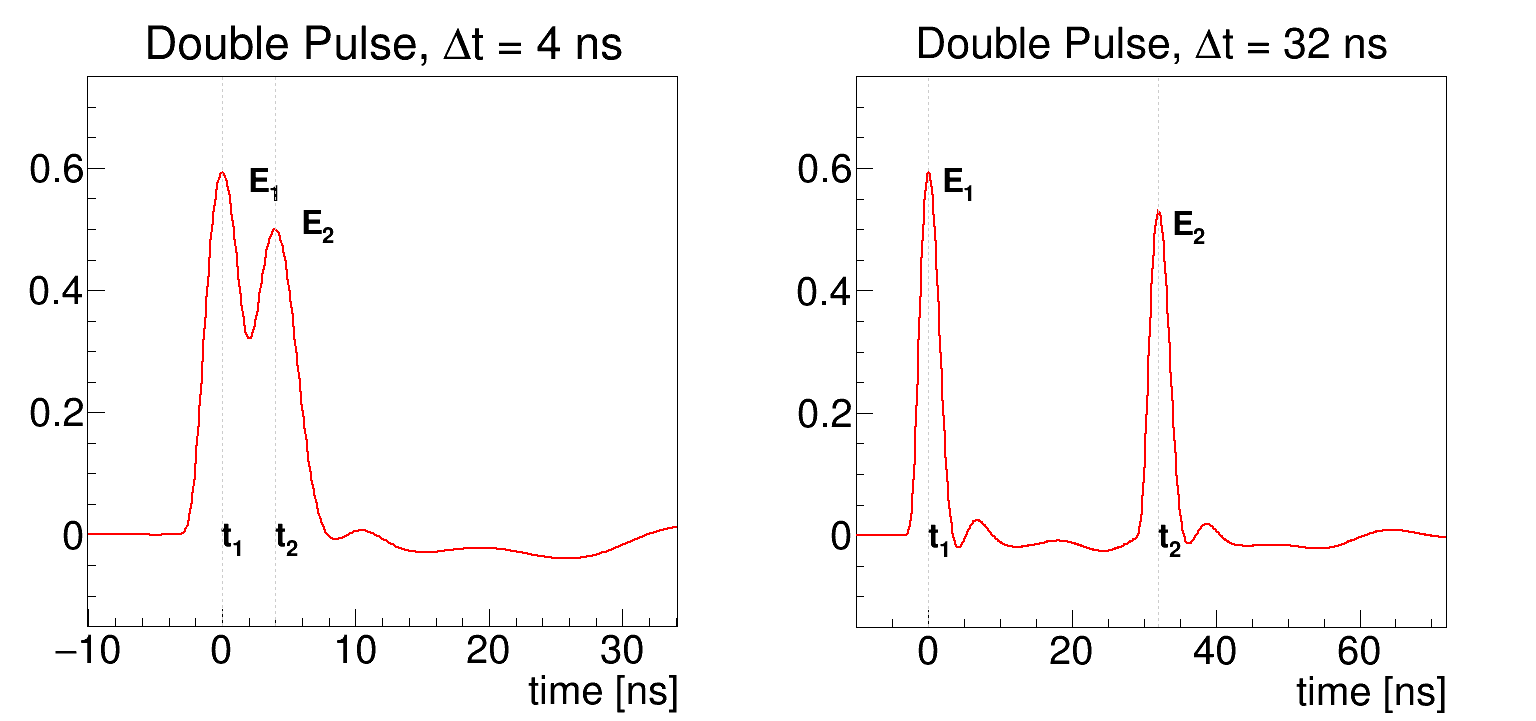}
\caption{Two typical double-pulse signals produced with different delays. The digitized data are fit to a pre-defined template.}
\label{fig:stdp}
\end{figure}

An external delay generator (DG) (SRS DG645) is used to send prompt and
delayed signals. The input to the DG is a replica of the Master Clock 
sent by the Laser Control Board. Two of the four DG outputs are
connected to the ODD and EVEN lasers, respectively. This enables the sending
to the same calorimeter of two pulses with a relative delay programmable
in the range $[0,1]$ sec in steps of $10$~psec. The ranges which
are relevant for the calorimeter response are (0-100)~ns, in steps
of $\sim 1$~ns, and $(0-100)\,\mu$s, in steps of $\sim
1\,\mu$s. 

%The DG allows also to send {\em bursts} of equally spaced pulses, thus
%mimicking the flash of particles which illuminate the calorimeters
%during beam injection.

%An additional feature of the DG, which turns out to be very useful in signal
%normalization, is the {\em prescale
  %option}: each DG output can be indipendently 
%prescaled by a factor $N=1-10000$, 
%such that only 1 signal every $N$ triggers is actually issued. 
%This feature is used to prescale the {\it prompt} signal. If $P$ is
%the prompt signal and $T$ is the delayed (or {\it test}) signal, 
%then by using the prescale it is possible to build the following
%sequence of pulses:
%$$
%P+T \,;\, T \,;\, P+T \,;\, T \,;\, ...
%$$
%
%This sequence allows to use the $T$-only signal as normalization.
%
%In the next sections, details on how the {\em Short Term} and {\em
  %Long  Term} Double Pulse are described.

\subsubsection{Short-Term Double Pulse}
\label{sssec:stdp}

The Short-Term ($20$~ns) data structure can easily be measured by
inserting the movable mirrors and by operating the delay generator, as
described in the previous section. The second pulse is delayed 
in the 0-80~ns range\footnote{The delay can be set to be much
  larger, but the Short-Term gain is back to 1 after 60-80~ns.}.

For each delay, a run is taken with few thousand events. 
%Half of the events
%consist in a prompt  and a test pulse, the so-called double pulse events,
%while the other half only have the test signal and are used as reference.
Two double-pulse events, from two different runs and with different delays, are
shown in Fig.\ref{fig:stdp}.

\begin{figure}[htbp]
\centering
\includegraphics[width=0.75\textwidth]{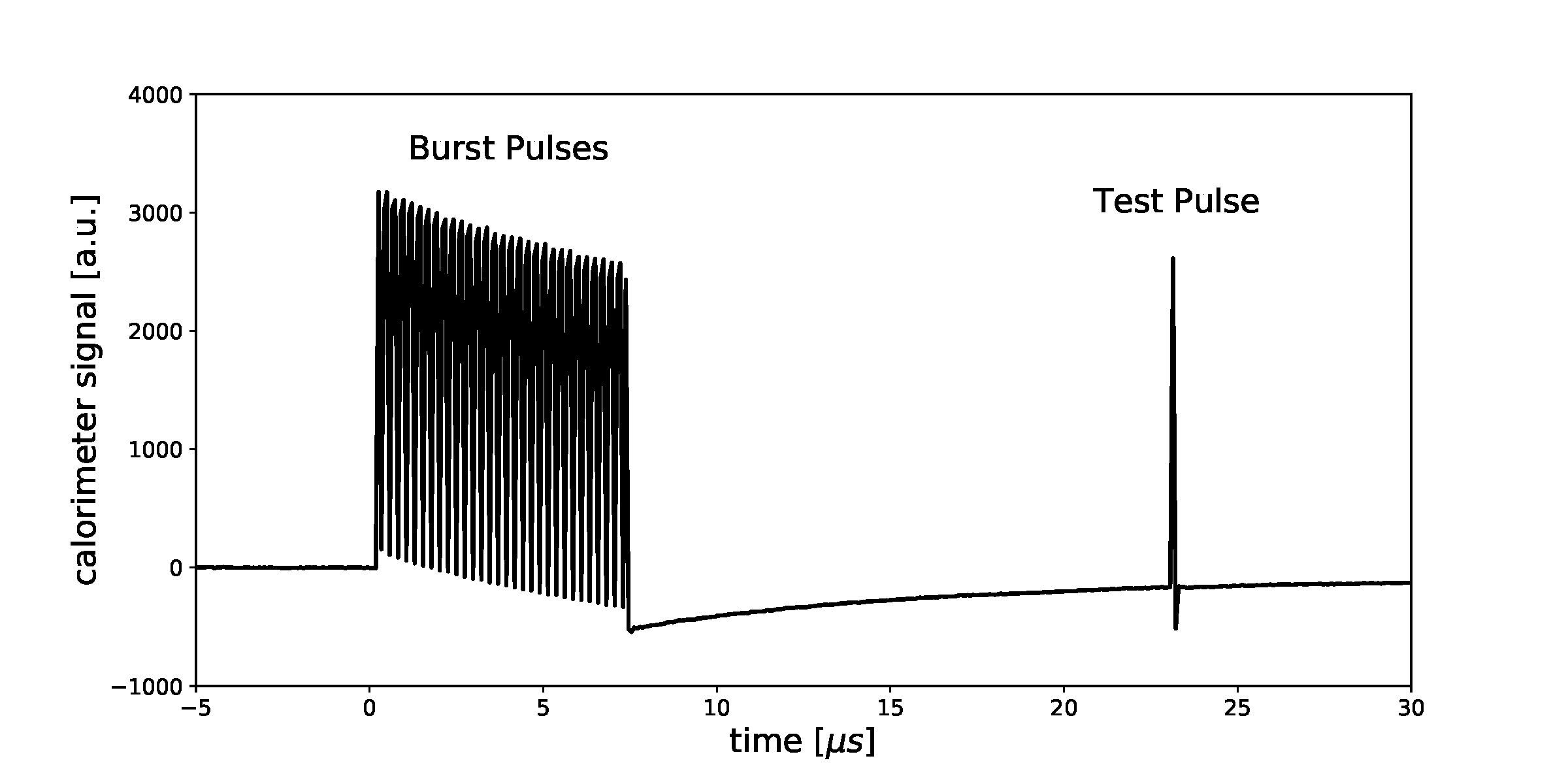}
\caption{Typical signal pattern for long-term double pulse. First, a burst of several, equally spaced, laser triggers is sent to simulate the arrival of multiple particles, and then a final probe laser pulse is sent with a large and variable delay with respect to the burst. Notice the signal baseline sag due to the initial burst of pulses.}
\label{fig:ltdp}
\end{figure}

\subsubsection{Long-Term Double Pulse}
\label{sssec:ltdp}

The longer time constant is more complex to measure. 
The gain drop is in fact due to the overlap of several pulses which
overload the HV power supply. Because of this, the prompt signal
is not  provided by a single pulse, but by a {\it burst} of pulses.

The test is performed as follows:
\begin{itemize}
\item the laser control board sends a burst of $N$ laser pulses separated
  by an interval $\Delta t$, normally set equal to $120\,$ns, which is the minimum delay allowed by the laser driver. The amplitude of these pulses can be modified, by
  using the filter wheel located after the laser head, by a factor $4-0.001$, the
  default value corresponding to an energy deposit of $\sim 1.5$
  GeV in a crystal.
  %The three parameter are set at run start, with
  %typical values $N$=50, $\Delta t$=120 ns, FW=4;
\item after a delay, programmable in the range 0-255 $\mu$s, 
  a test pulse is sent from the second laser.  
\end{itemize}

An example of such a pulse structure is shown in Fig.~\ref{fig:ltdp}.

The trigger for the delayed test pulse is provided by the Laser Control
Board, described in Sec.~\ref{ssec:control}.

\subsection{Gain calibration} 
\label{ssec:absolute}

\begin{figure}[htbp]
\centering
\includegraphics[width=0.5\textwidth]{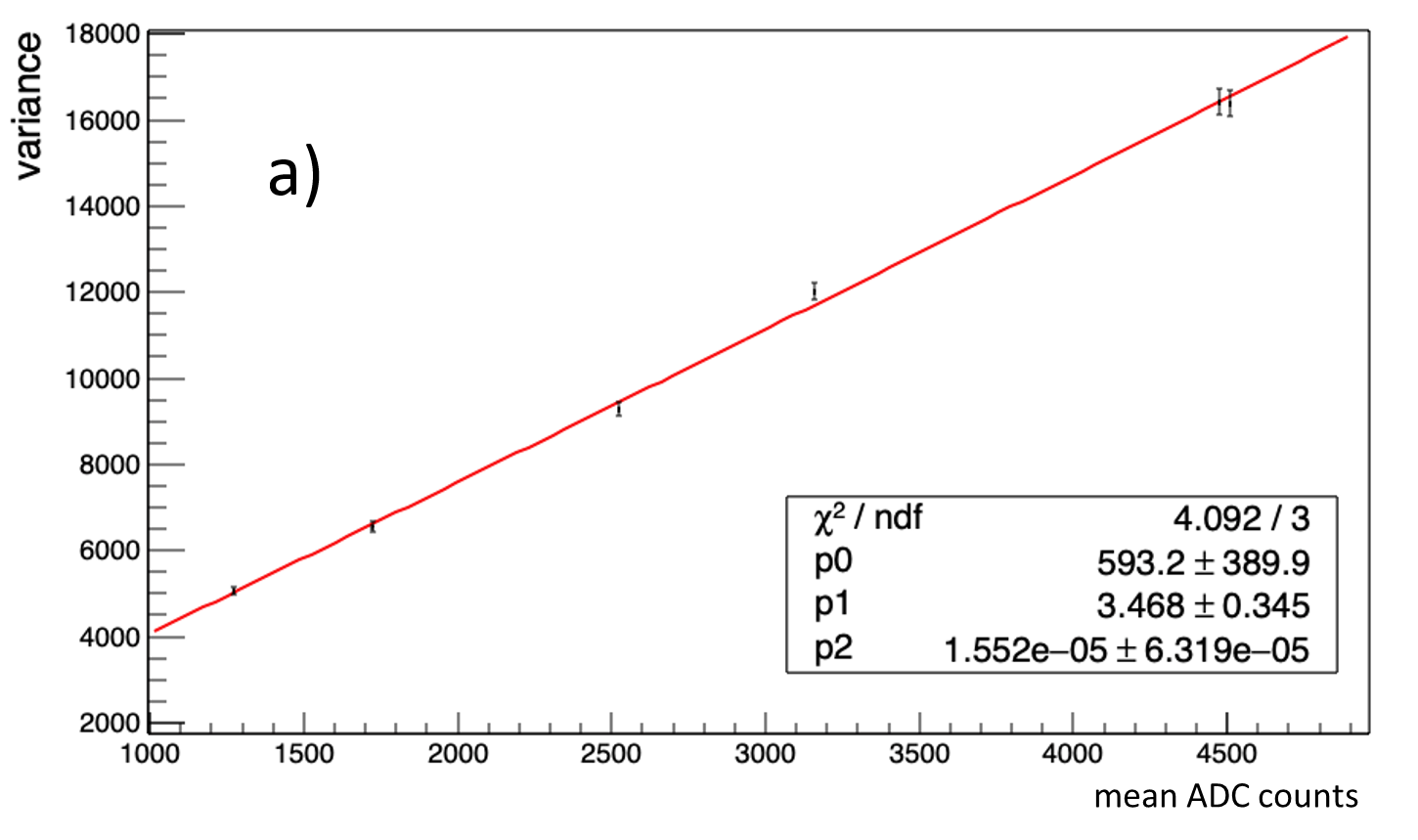}
\includegraphics[width=0.45\textwidth]{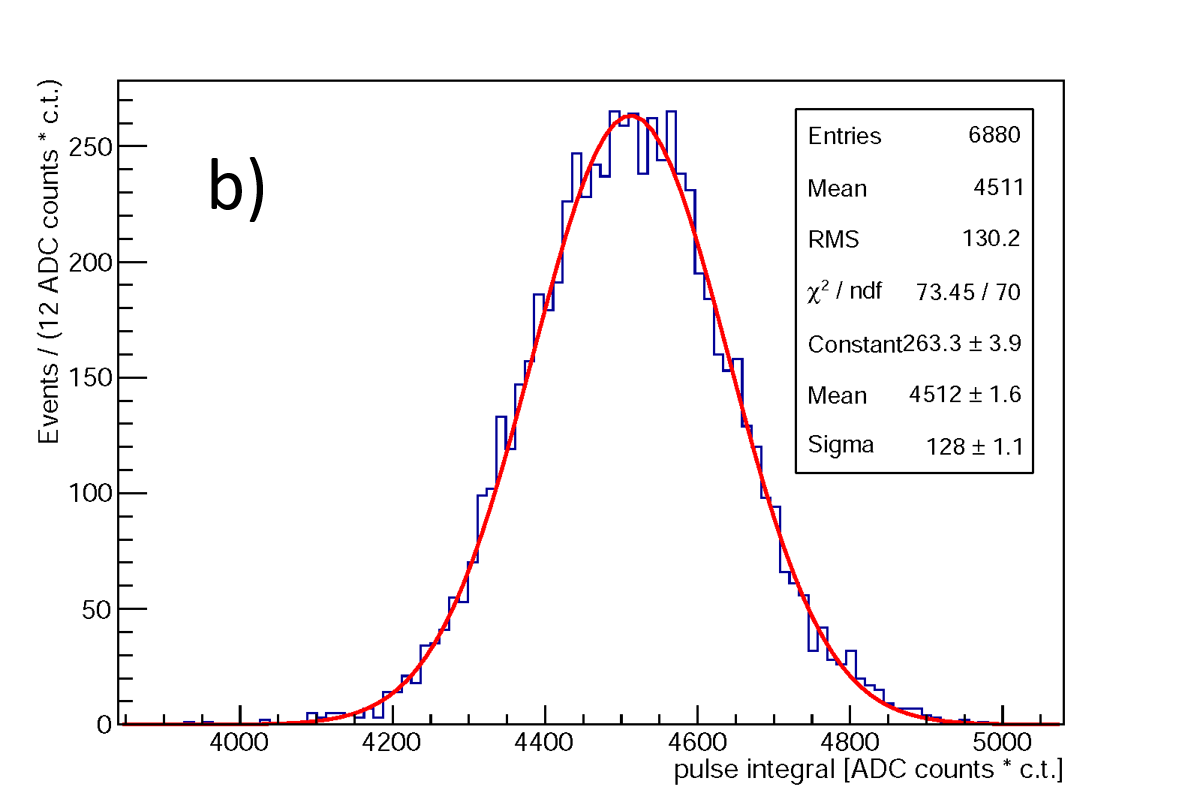} %{FW6_bis.png}
\caption{In a) plot of variance $\sigma^2$ versus mean pulse-integral, $M$ of a distribution of fitted laser pulses on a SiPM. The different discrete mean values are obtained using a multi-step filter wheel to attenuate the light intensity.  
%The sequence of runs begins and ends in the open position of the filter wheel, corresponding to the upper right most point. 
The inverse of the fitted slope corresponds to the $G_{\rm pe}/M$ and the good linearity implies that the variance depends only on the statistics of the number of pixels fired on each event. In b) a typical distribution of pulse energies for a given laser intensity.}
\label{fig:gaincal}
\end{figure}

The laser system is used also to equalize the gains of the 1296 SiPMs in the 24 calorimeters around the $g-2$ ring. This procedure, described in detail in Ref.~\cite{Fienberg2015}, is used to extract, for each SiPM at a given temperature and bias voltage, the calibration constant $G_{\rm pe}=M/N_{\rm pe}$ relating the number of photo-electrons $N_{\rm pe}$ generated in the SiPM to the mean pulse-integral, $M$. This value is obtained by recording thousands of laser pulses, fired at constant intensity and repetition rate. The distribution of the integrals of the recorded pulse signals returns a mean $M$ and a variance $\sigma^2$. By varying the laser intensity, for example by using the calibrated filter wheel in front of each laser head, a graph of the variance $\sigma^2$ as a function of $M$ is obtained, showing a linear dependence (see for example Fig.~\ref{fig:gaincal} and Ref.~\cite{Kaspar2017}). 

Assuming  the distribution of photons from the laser source to be Poissonian and the mean $M$ proportional to the number of photo-electrons $N_{\rm pe}$ through a gain constant $G_{\rm pe}$, the slope of the variance/mean graph gives the proportionality constant $G_{\rm pe}$ directly. 
This calibration constant is different from one SiPM to another and depends on both temperature and bias voltage, in particular on the SiPM {\it over-voltage}, the difference between the bias and the breakdown voltage. The gain constants can vary in time due to changes in environmental conditions and the photo-electron   
calibration is thus used when required to equalize the response of all SiPMs in a calorimeter. 
%This is done primarily by adjusting the programmable gain amplifiers on the SiPM readout boards and, in the end, by finely tuning the over-voltages.
%In a next step, not involving the laser system, the number of photo-electrons is calibrated with respect to the energy lost by the positrons inside the PbF$_2$ crystal.
While the laser system is used to measure the gain of the SiPMs, the laser
system is not involved in the next step of the process used to determine the
number of photo-electrons produced per unit energy lost by the positrons
inside the PbF2 crystals.

\subsection{Flight Simulation Mode}  
\label{ssec:flight} 
%(....) (Anna Driutti; Udine)

\begin{figure}[htbp]
\centering
\includegraphics[width=0.75\textwidth]{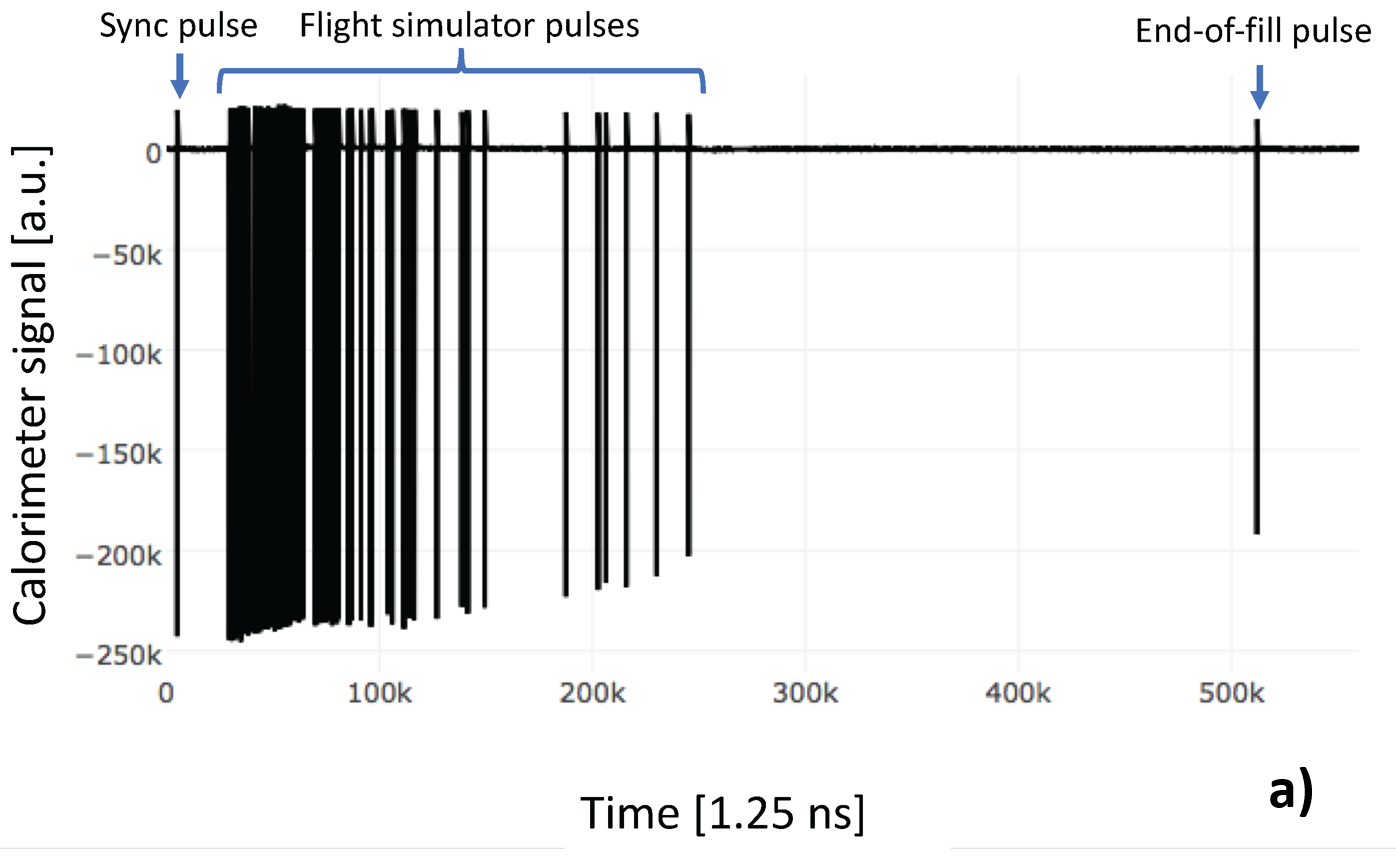}
\includegraphics[width=0.75\textwidth]{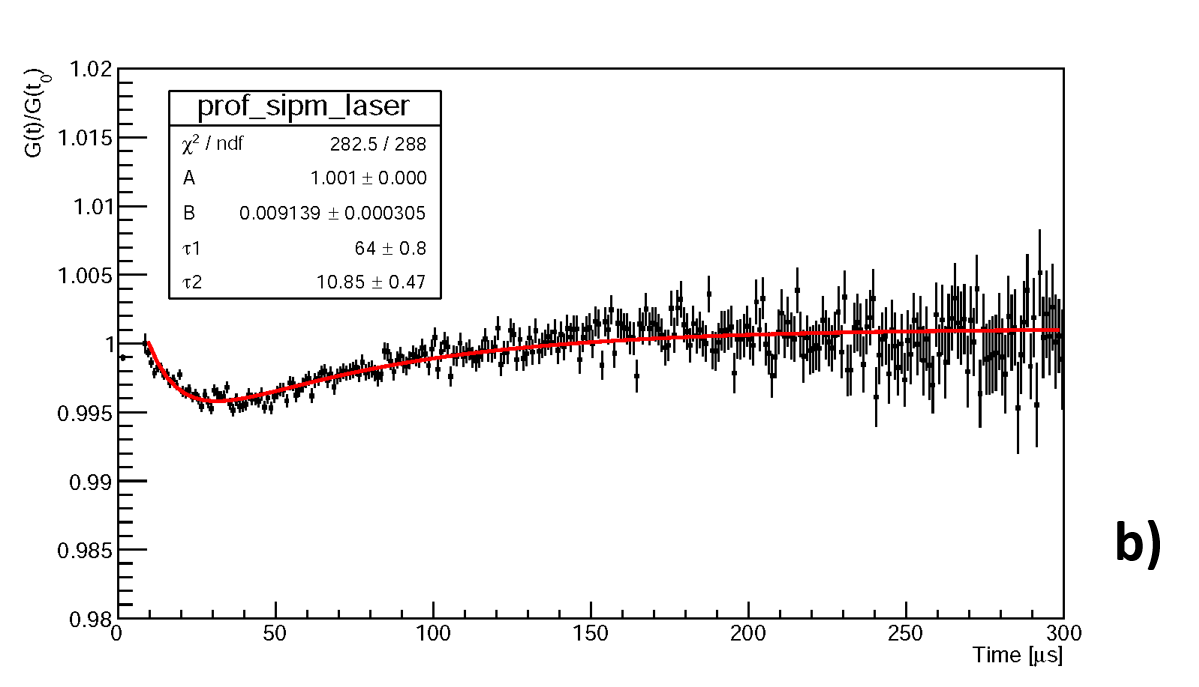}
\caption{a) Example of the distribution of laser pulses inside a fill when the system is in Flight-Simulator Mode with $N_p = 64$. On the $x-$axis the time is expressed in clock ticks (a clock tick corresponds to 1.25 ns). The first FSM pulse arrives 8 $\mu$s after the synchronization pulse. b)  Typical gain function obtained when the laser fires in FSM, fitted using Eq.~\ref{gainfit}. The legend details the fit results.}
\label{fig:FSM}
\end{figure}

As already stated in Sec.~\ref{sec:intro}, the gain of the SiPMs is affected by the high rate of the decay positrons in a fill. This effect, if not accounted for, could lead to a significant systematic uncertainty in the determination of $\omega_a$. An important feature of the Muon $g-2$ laser calibration system is that it can simulate these In-Fill gain variations, $\Delta G(t)=G(t)/G_0$ where $G_0$ is the OoFPs reference gain, using the Laser Control Board in the Flight Simulation Mode. The FSM, although not used explicitly to compute the gain correction tables of Sec.~\ref{ssec:performances}, has shown the importance of the In-Fill Gain variations, as illustrated in Fig.~\ref{fig:FSM}b. The FSM has been also extensively used to assess the calorimeters and DAQ performance during the commissioning phase.

As described in Sec.~\ref{ssec:control}, the LCB in FSM triggers the laser with a sequence of pulses that simulates the rate of decay positrons within a fill, i.e according to an exponential distribution.
In addition, the LCB allows for the selection of the average number of calorimeter's hits $N_p$ expected per fill  and it is possible to change the intensity of laser light using the filter wheels positioned on the optical table. 
It should be stressed that the anomalous frequency $\omega_{a}$, present in the real data,  is not introduced, i.e. no wiggle appears in the distribution of arrival times of the simulated positrons. Moreover, there is a certain difference with a real muon fill because all the 1296 crystals are simultaneously illuminated with laser pulses, while in the case of a real positron from the muon decay only a few channels per calorimeter are activated. This means that the SiPMs power supplies, which also influence the gain drop, are stressed more in the flight simulator mode than in real data taking.
An example of the distribution of laser pulses in  FSM with 64 simulated 1.8~GeV-positrons inside one fill is shown in Fig.~\ref{fig:FSM}a.
Fig.~\ref{fig:FSM}b shows a typical gain drop obtained with a 96-hit flight simulator and a filter with 100\% transmittance. It is fitted using a functional form that accounts for the time constant of the muon lifetime, $\tau_1$, and the time constant of the SiPM recovery time, $\tau_2$, according to the equation:

\begin{equation}
\label{gainfit}
\frac{G(t)}{G_0} = A - B \left( e^{-t/\tau_1} - e^{-t/\tau_2} \right)
\end{equation}

where $A$ (expected to be equal to 1) 
is the gain ratio %for $t=t_0$
at the starting time of the pulse bunch, 
%which is  expected to be equal to 1,
$B$ is a factor quantifying the gain drop and  $\tau_1$ and $\tau_2$ are the two time constants. 

The LCB can be used also for simulating the initial splash, i.e. the arrival of many particles, mostly muons that are not captured by the ring inflector, on the calorimeters close to the beam line. This is done by pulsing the laser 100 times at the beginning of the fill, before using the FSM to simulate the decay positrons detected in the fill. This feature has been used during the commissioning phase, starting from the moment the muon beam was present for the first time and the importance of the splash effect was recognized.
%This feature has also been extensively used during the commissioning phase. 
%Figure~\ref{fig:FS+S} shows and  (a) example of the sequence and (b) an example of the gain function obtained with this mode. 

 %\begin{figure}[htbp]
 %\captionsetup{labelsep=period}
	%\begin{center}
%\subfloat[]{\includegraphics[width=0.5\textwidth]{fs_w_splash}}
%\subfloat[]{\includegraphics[width=0.5\textwidth]{run13776_flight_sim_wsplash_and_pfx.png}}	\\
%\end{center}		
 %\caption{Figure (a) shows an example of pattern pulses used to simulate the positron decay during the fill and the initial splash after the beam injection, and figure (b) shows the resulting gain sag. The box in the top of figure (a) contains a blow up of the first $50\,\mu s$ of the fill to detail the simulated splash.}
	%\label{fig:FS+S}
%\end{figure}

\subsection{Assessment of laser performance and gain calibration procedure}
\label{ssec:performances}

Ultimately, the laser system is leveraged to monitor and correct gain fluctuations in the calorimetry system.  The aggregate correction is classified into different types of corrections.  The long-term drift correction is referred to as the Out-of-Fill Gain (OoFG) correction.  The systematic shift due to effects correlated with muon injection is called the In-Fill Gain (IFG) correction.  The nanosecond timescale effects of one pulse on the subsequent pulse is the short-time double pulse (STDP).   All these corrections are performed on each individual SiPM detector of the 1296 total.

\subsubsection{Laser monitor performance}
\label{sssec:monitor}

\begin{figure}[htbp]
\centering
\includegraphics[width=0.45\textwidth]{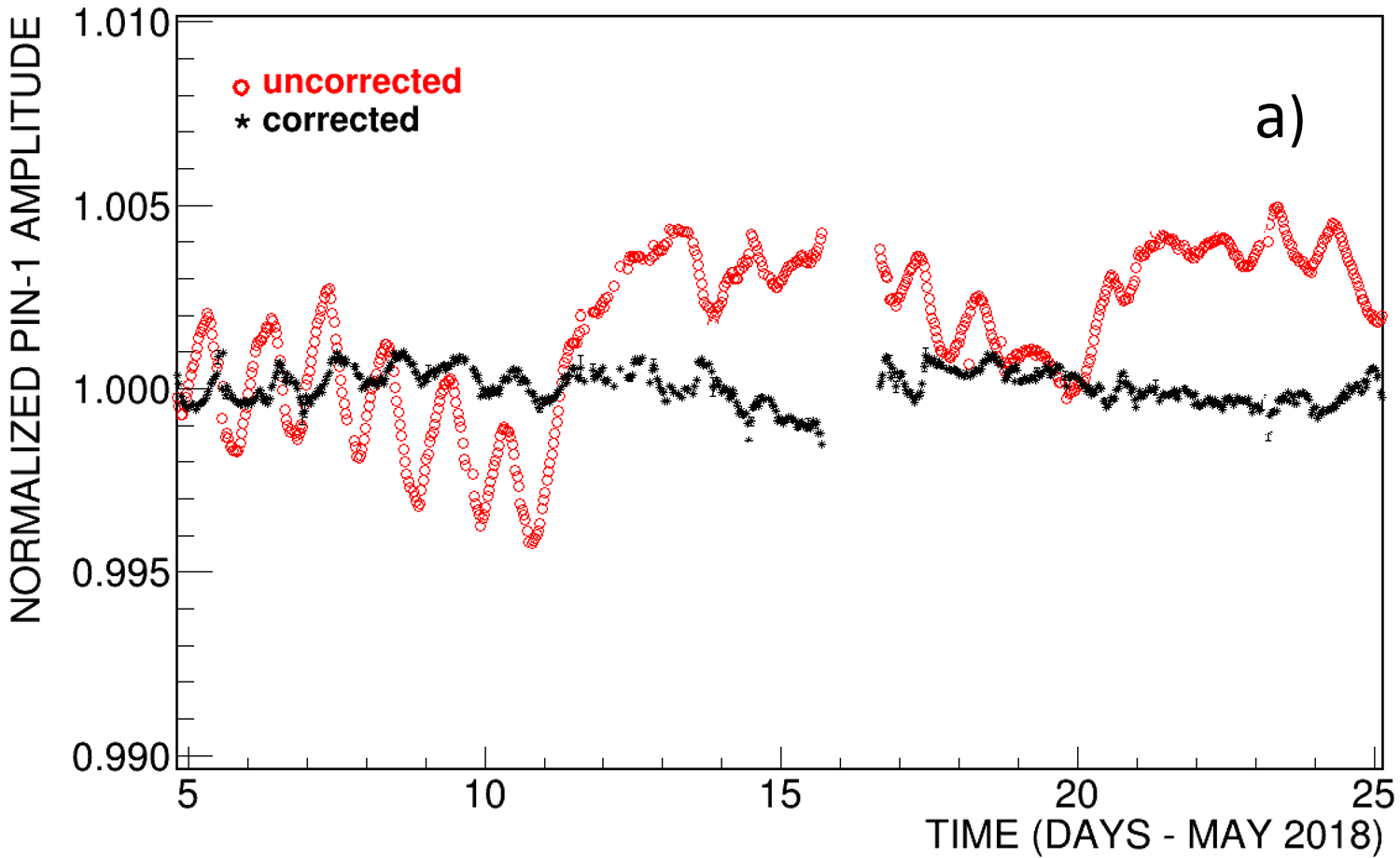}
\includegraphics[width=0.45\textwidth]{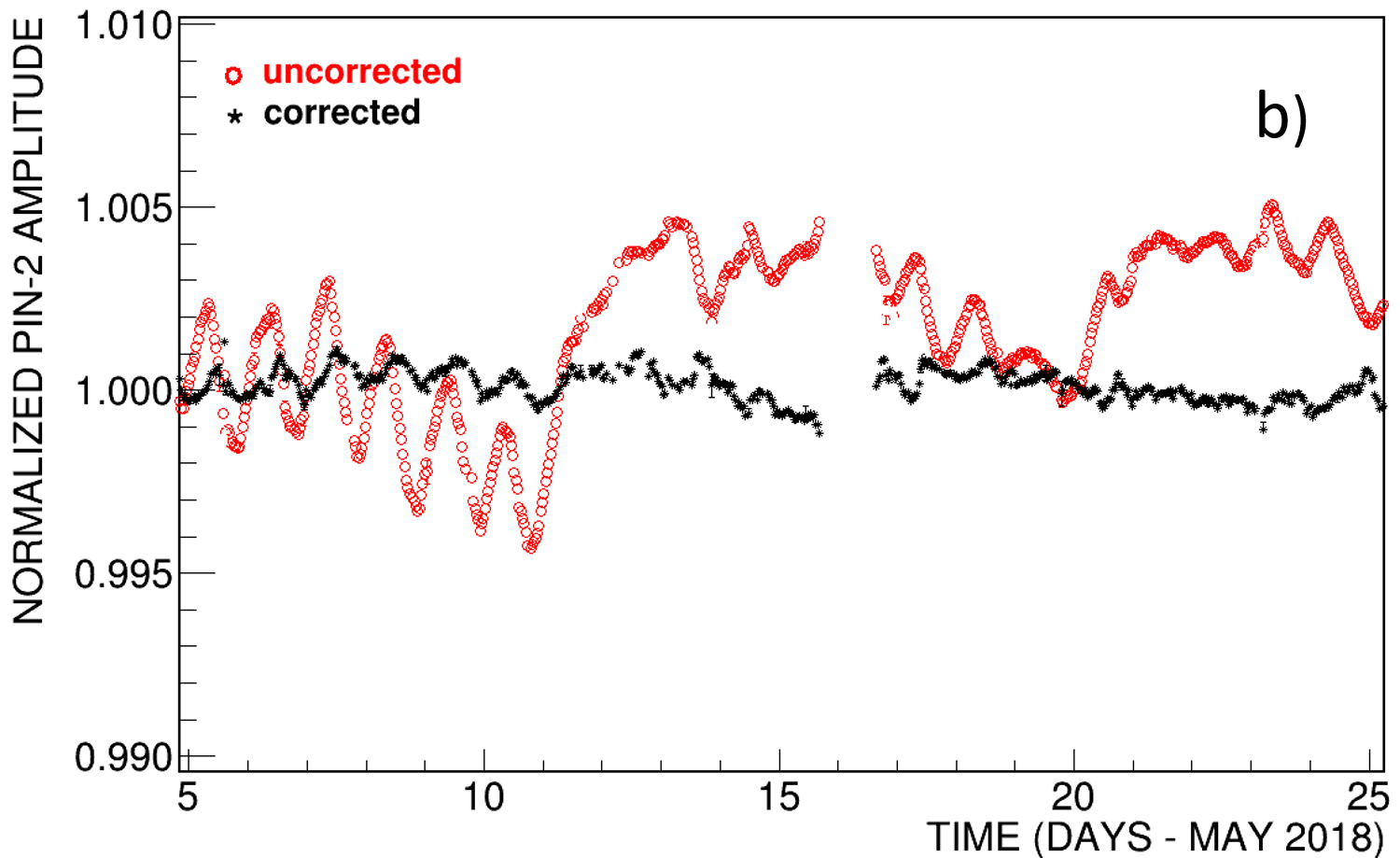}
\includegraphics[width=0.45\textwidth]{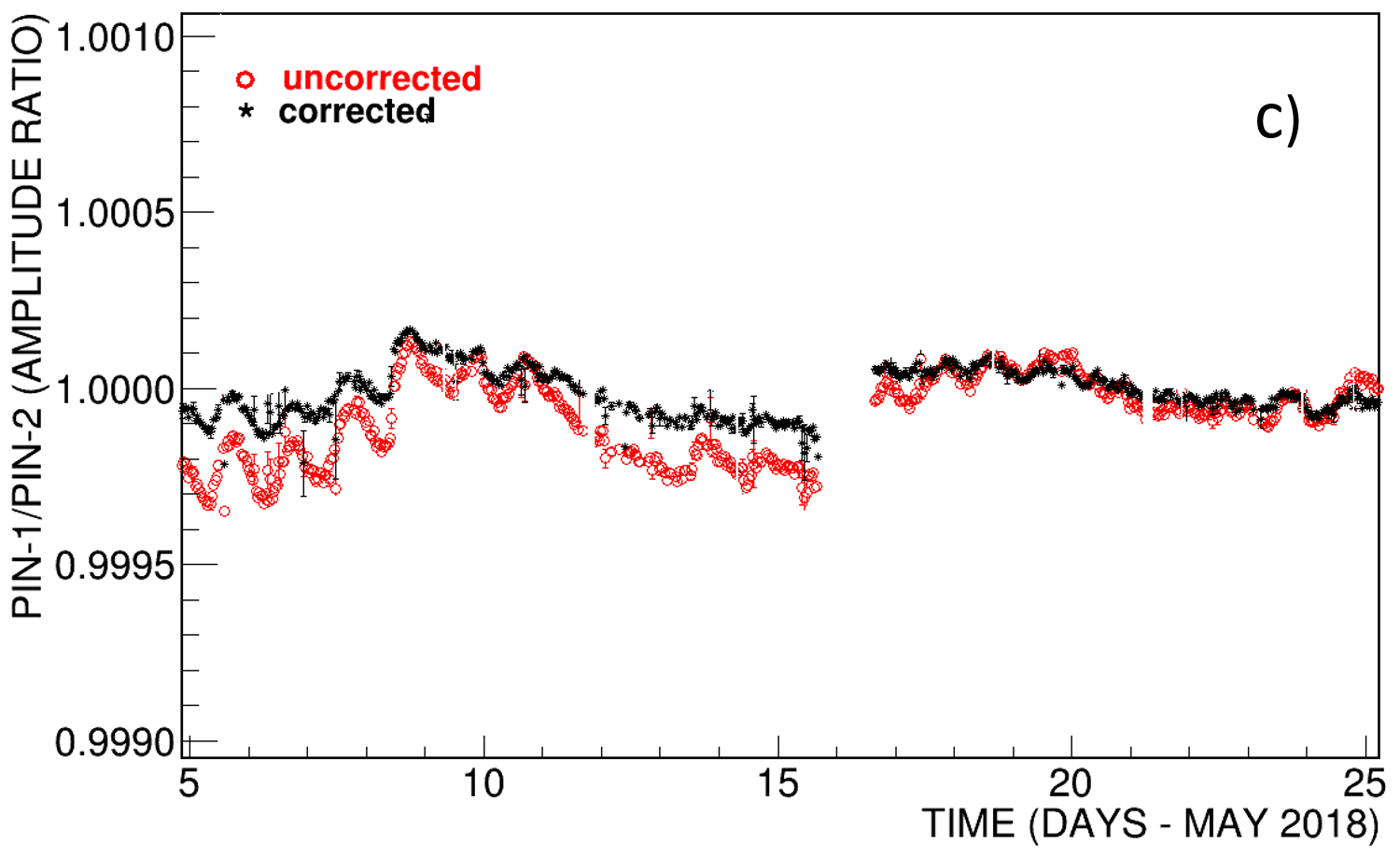}
\includegraphics[width=0.45\textwidth]{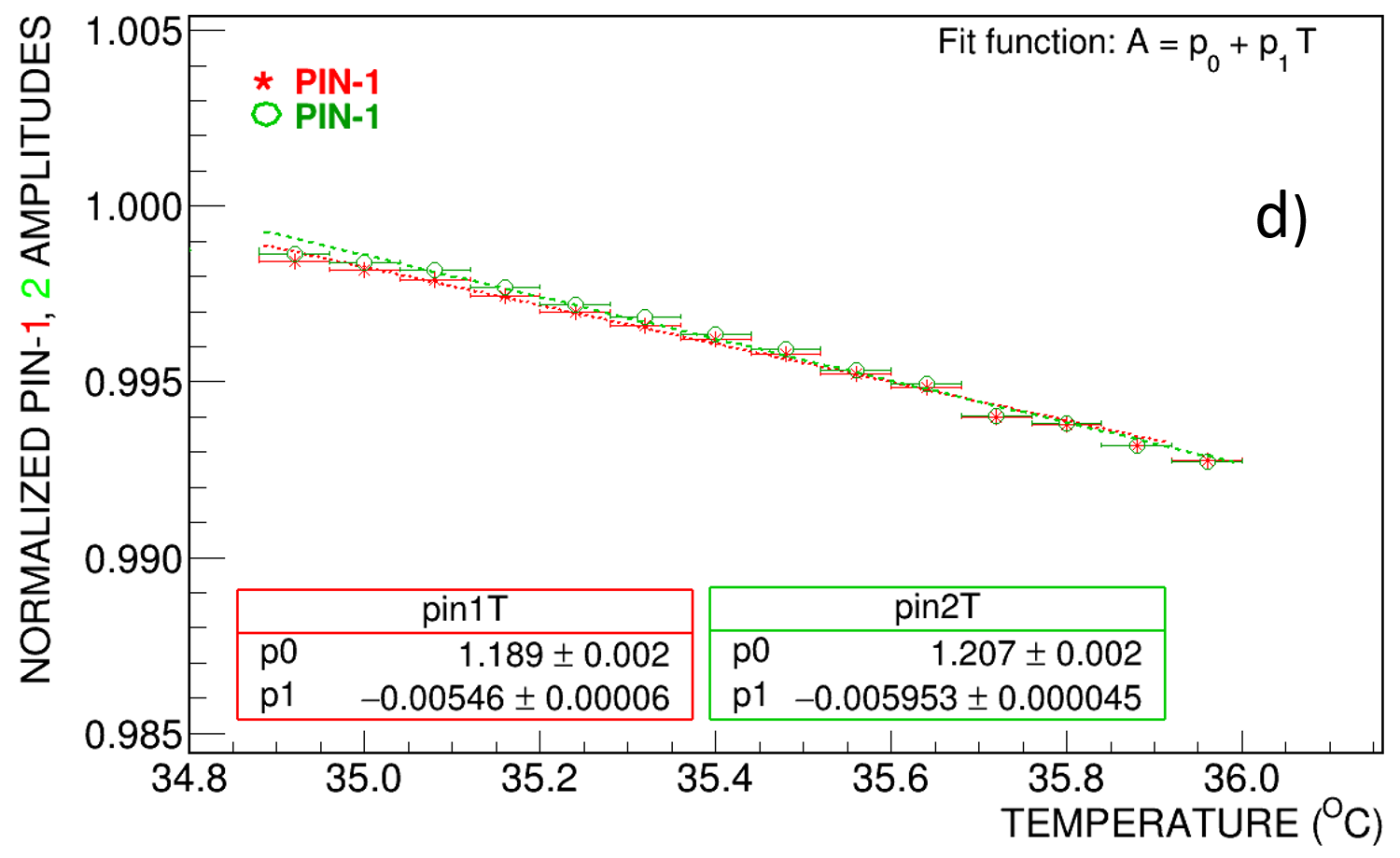}
\caption{Plots assessing the stability of the Source Monitor versus temperature changes in a two-week period. Upper panel: (a) and (b)  the SM PIN1 and PIN2 signals, respectively, before and after temperature corrections. Lower panel: in (c) PIN1 to PIN2 ratio before and after temperature corrections; in (d) PINs temperature dependence for the same data set. The relation between the two quantities is almost linear, therefore a linear correction is applied to the data to compensate for this effect. The final stability is better than $0.2\%$ for the single PIN and $10^{-4}$ for the ratio.}
\label{fig:SM_perf}
\end{figure}

The performance of both Source and Local monitoring systems has been assessed with real data. The SM response is temperature dependent. The temperature of each of the 6 SMs is measured directly by the Monitoring Boards and its effect can thus be accounted for. As shown in Figs.~\ref{fig:SM_perf}a-b, the fluctuations of the response of each individual PIN of the same SM are within $\pm 5\times 10^{-3}$ in a 3-weeks period but reduce to $\pm 10^{-3}$ when temperature corrections are implemented. The ratio between the responses of the 2 PINs, Fig.~\ref{fig:SM_perf}c, is in the $\pm 10^{-4}$ range, once temperature corrections are implemented.

Fig.~\ref{fig:SM_perf}d shows the dependence of the single PIN response on the temperature, for the same given SM. An almost linear  dependence of the PIN response is observed, with a negative $0.4\%$ variation per degree Celsius. In the corrected data only the linear correction is applied.

\begin{figure}[htbp]
\centering
\includegraphics[width=0.45\textwidth]{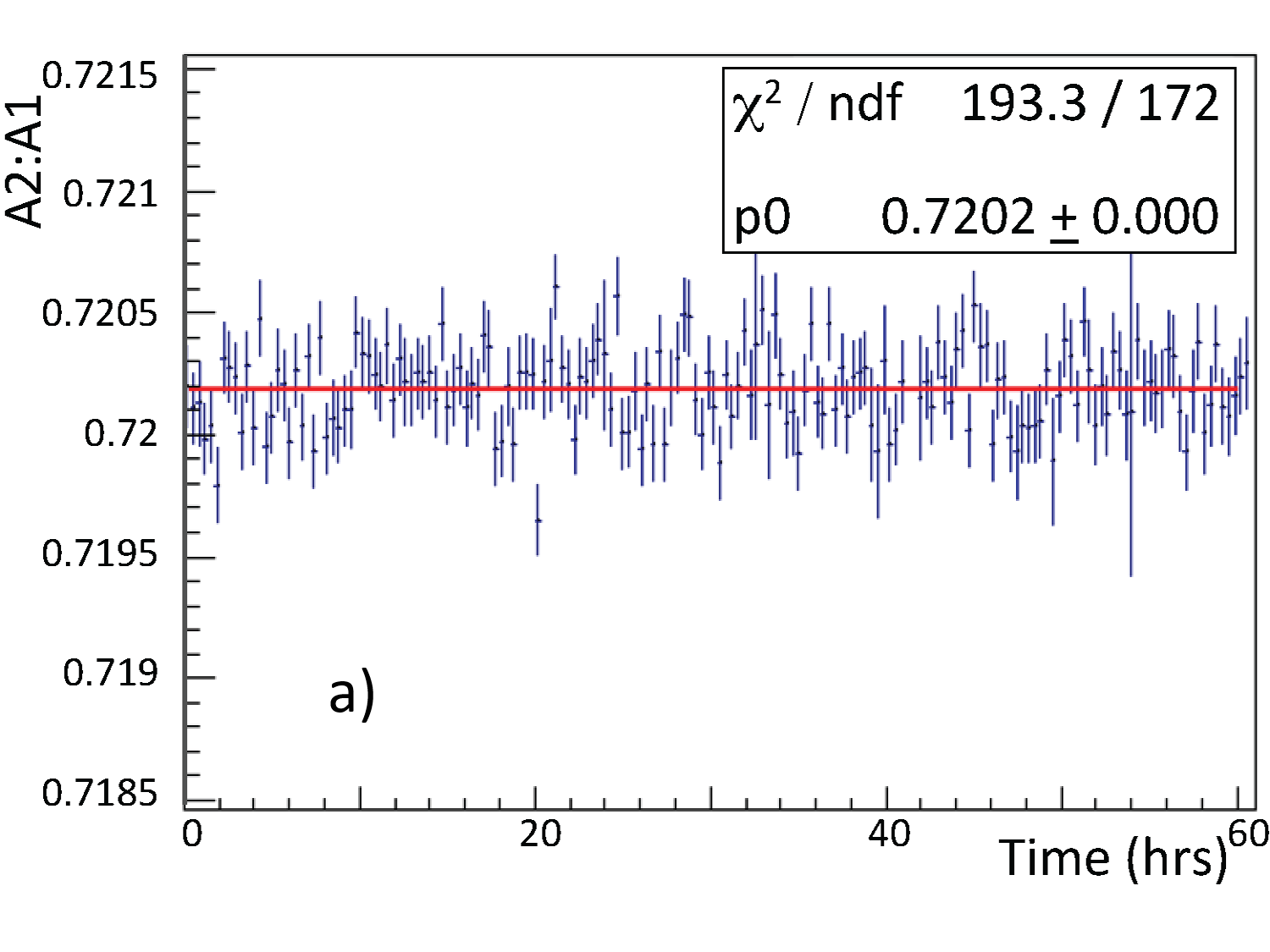}
\includegraphics[width=0.45\textwidth]{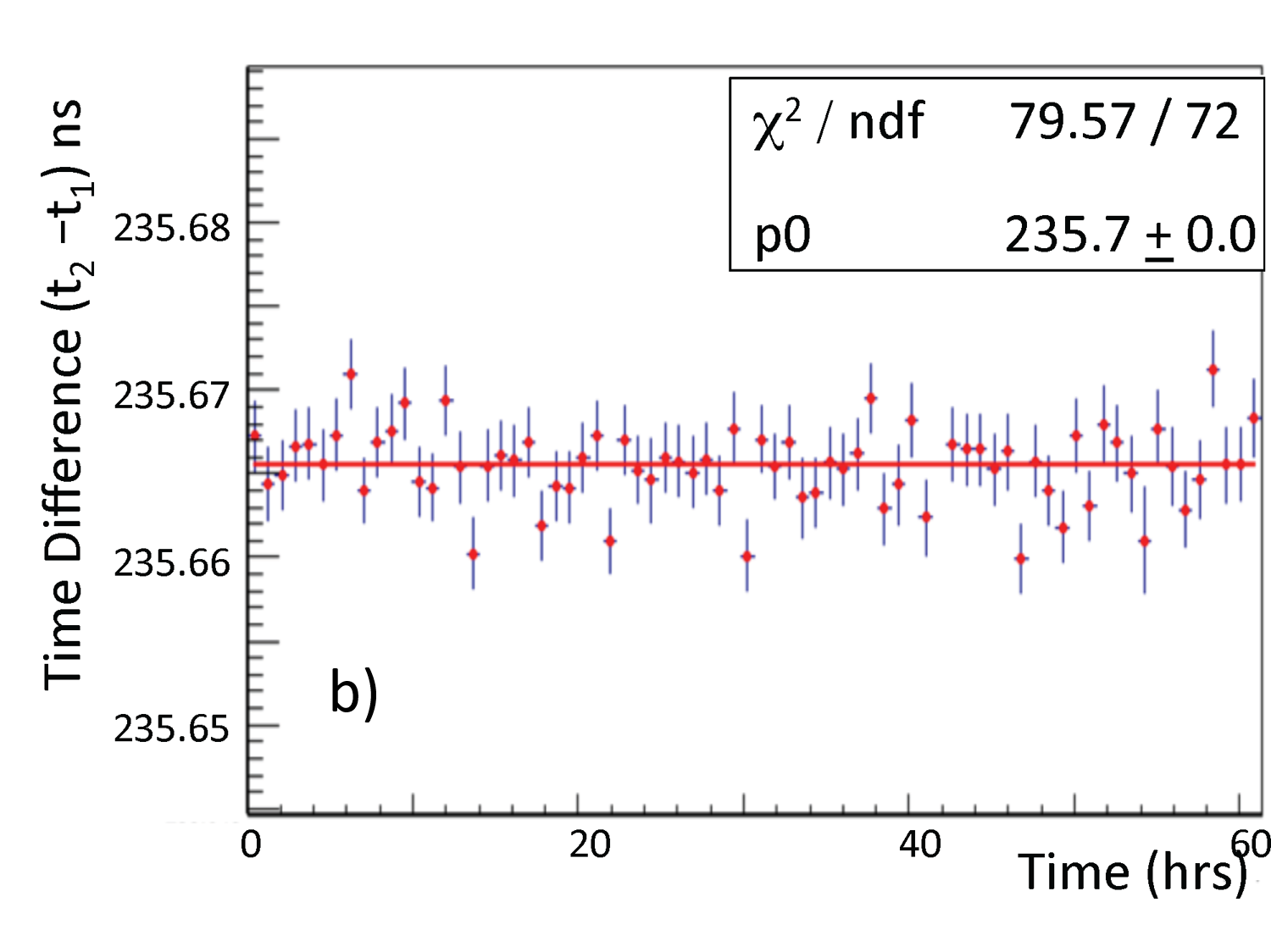}
\caption{Performance of the Local Monitor: in (a) ratio of amplitudes of the delayed LM peak to the respective SM peak for one particular LM  channel and, in (b), arrival time differences between delayed LM and SM pulses for the same channel.}
\label{fig:LM_perf}
\end{figure}

Figs.~\ref{fig:LM_perf}a-b show the stability of the LM, measured by the ratio of the LM delayed pulse to the SM reference one, and the time difference between the two pulses. In this case the expected stability of the ratio is lower than that of the SM, as it contains the fluctuations of the SM itself. Nevertheless it is below the per-mille level.

\subsubsection{Out-of-Fill Gain correction}
\label{sssec:out-of-fill}

\begin{figure}[htbp]
\centering
\includegraphics[width=0.45\textwidth]{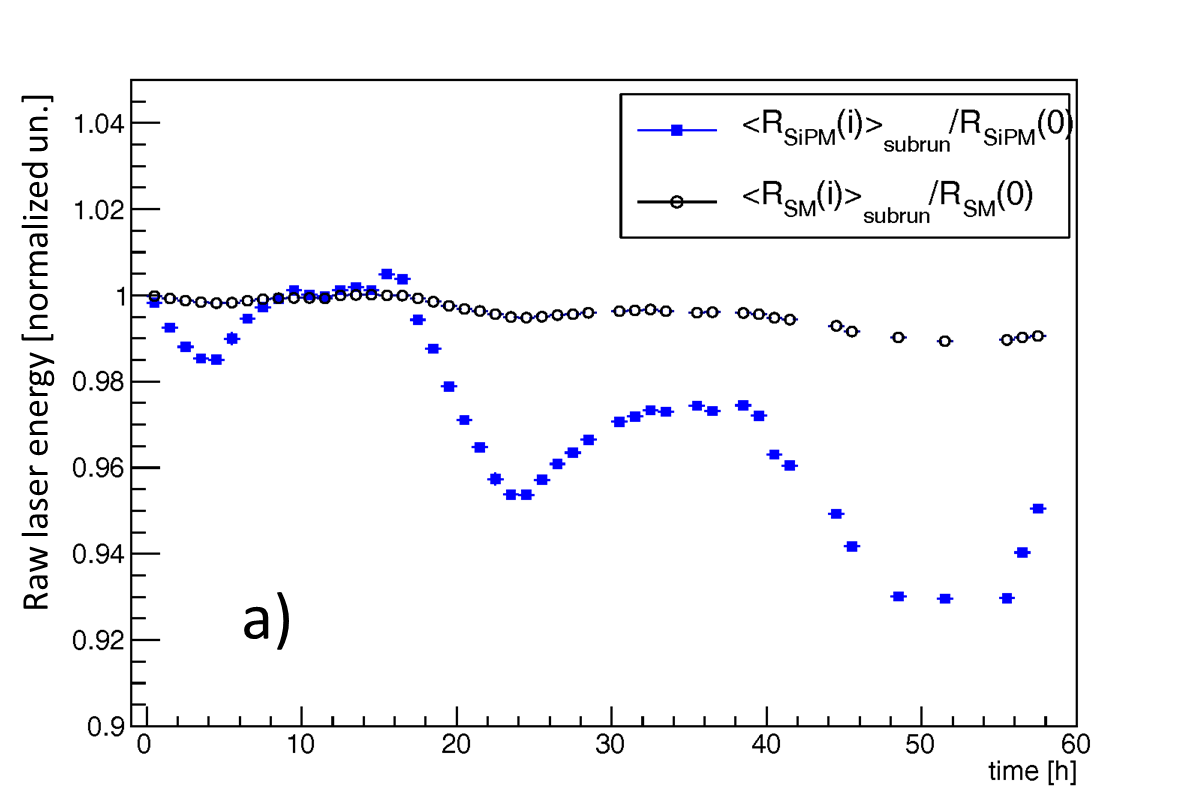}
\includegraphics[width=0.45\textwidth]{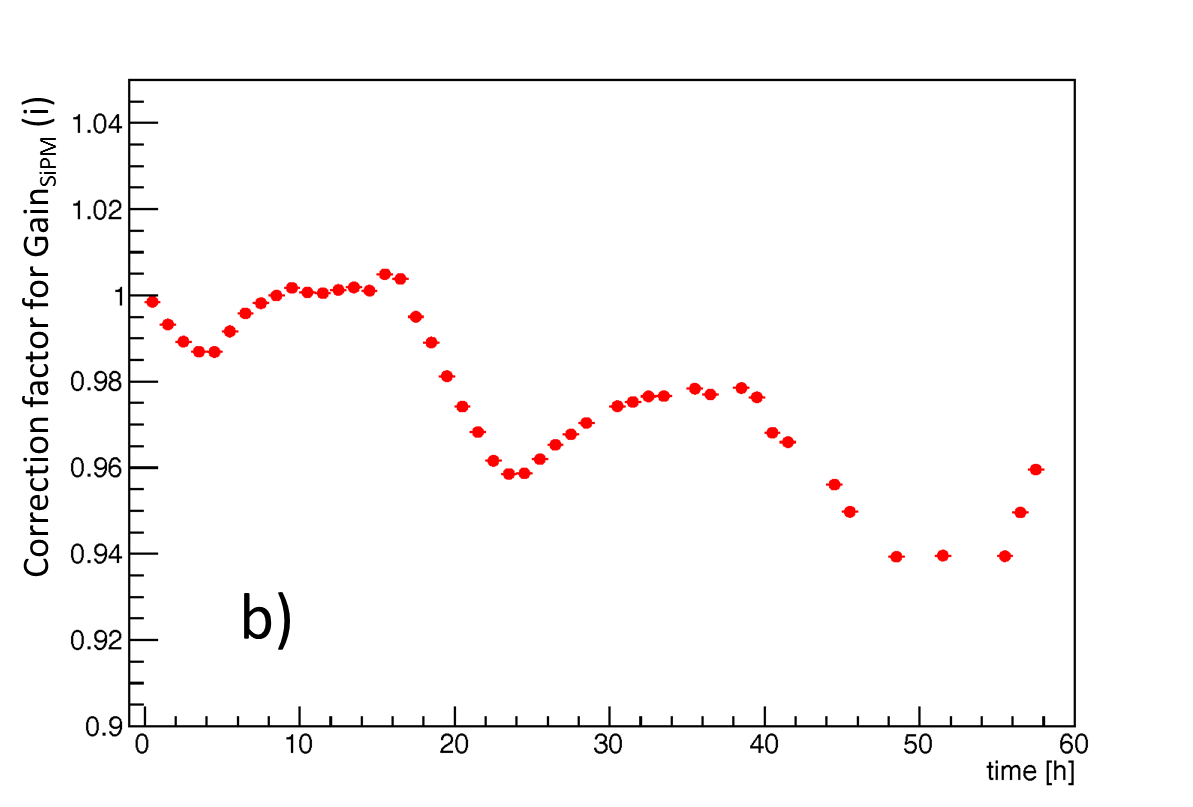}
\includegraphics[width=0.45\textwidth]{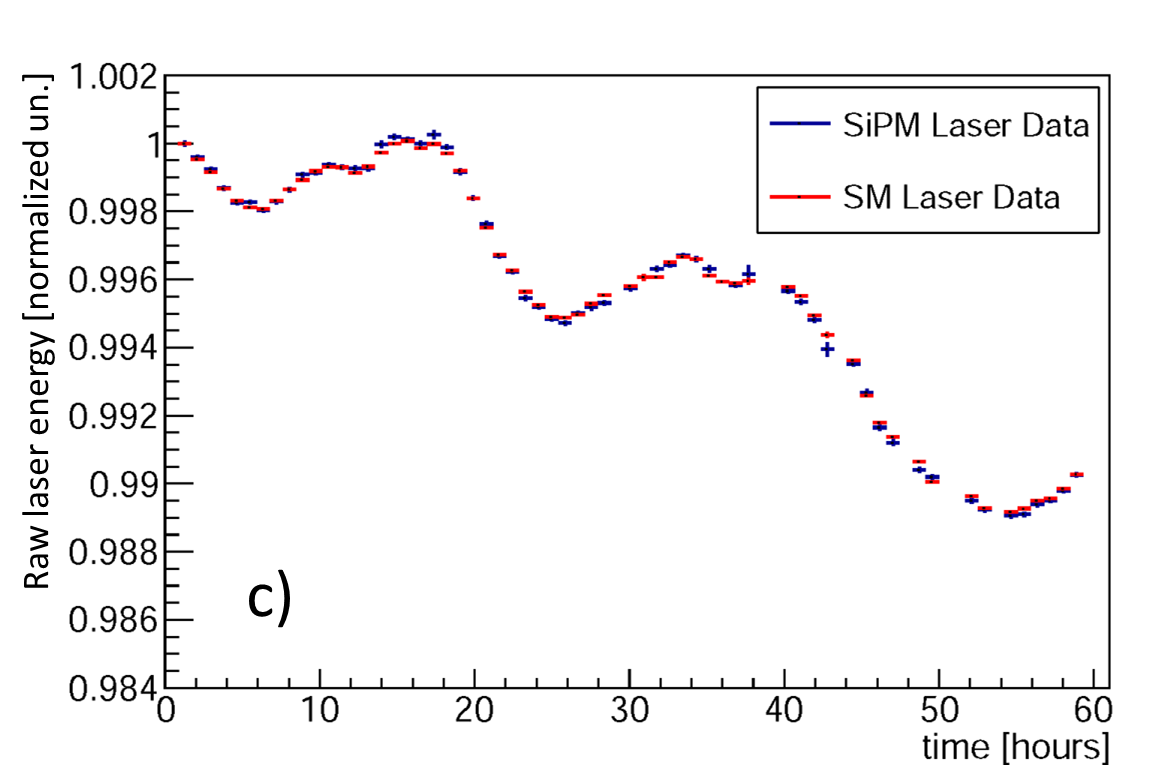}
\includegraphics[width=0.45\textwidth]{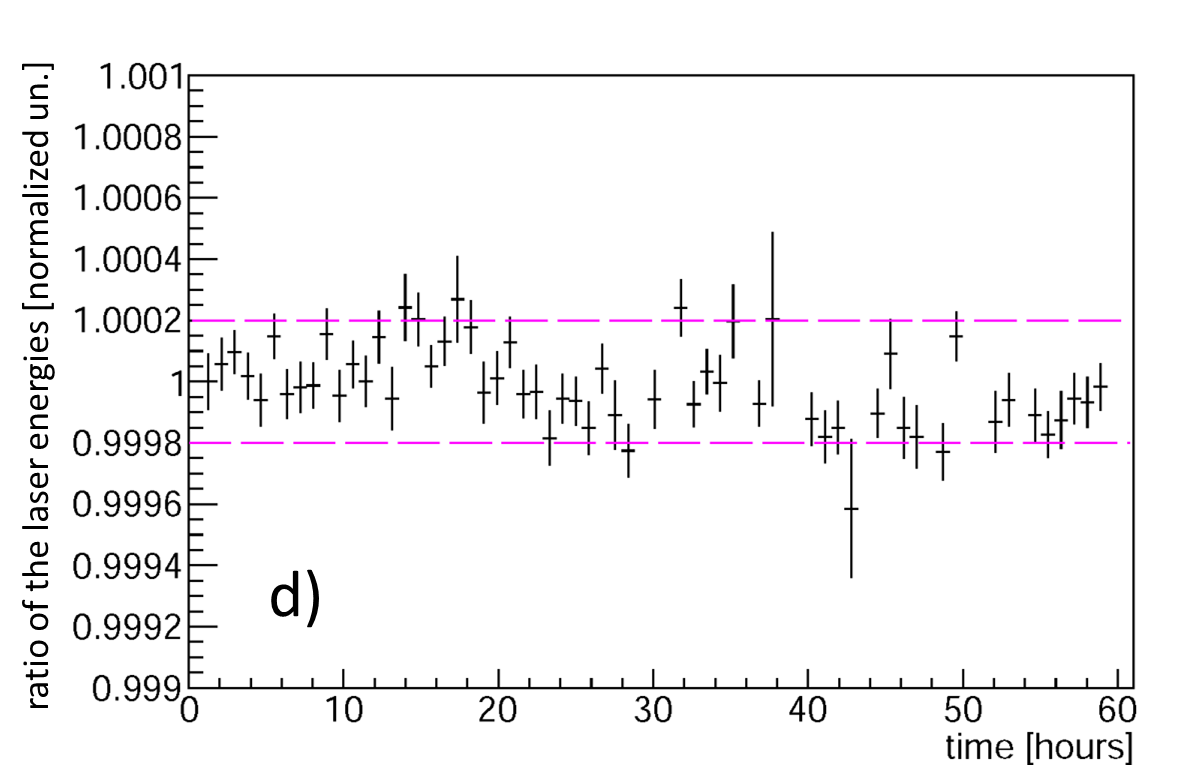}
\caption{The principle of the Out-of-Fill gain correction: in (a) the subrun-averaged raw energies of one SiPM (OoF pulses) and of the SM energies are used to compute the correction factors (b) for the SiPM gain. In (c) the OoFG-corrected SiPM energies (IF pulses) are compared to the SM energies: their ratio (d) is stable to $\pm1\times 10^{-4}$ on the long timescale, showing the consistency of the OoFG correction procedure. In this particular plot the stability exceeds the experimental requirements.}
\label{fig:oofg}
\end{figure}

Procedurally, the OoFG correction is calculated first and applied to all data, positron and laser.  Between the muon fills, OoF laser pulses are sent to monitor the response of the system.  Without the fill-dependent effects in the hardware, the response of the calorimeter system to laser pulses is the current state of the ever drifting detector gain.  
Temperature-related effects are the primary cause of long timescale, i.e. seconds, drift in the gain of the system. The OoFG correction folds all these effects and accounts for the aggregate effect.
%The primary cause of long timescale, i.e. seconds, drift in the gain of the system is due to temperature-related effects.  
Temperature is known to affect the amplification of SiPMs, PMTs, PIN diodes, the intensity of the lasers, and the transmission of fiber optic cables at different levels.  Figure~\ref{fig:oofg}a shows the response of one of the calorimeters' SiPM and the SM PIN diode.  An implementation of the correction $G_{\rm SiPM}$ for each SiPM, shown in Fig.~\ref{fig:oofg}b, is given by the following equation:

\begin{equation}
G_{\rm SiPM}(i) = \frac{\left< R_{\rm SiPM}(i) \right>_{\rm subrun}}{R_{\rm SiPM}(0)} \\
\cdot \frac{R_{\rm SM}(0)}{\left< R_{\rm SM}(i) \right>_{\rm subrun}},
\end{equation}

where $R_{\rm SiPM}$ is the response amplitude of the SiPM to the laser, $R_{\rm SM}$ is the response amplitude of the Source Monitor to the laser, the bracket variables indicate averaging all data over a sub-run (about 5 seconds), and $R_{\rm SiPM}(0)$ and $R_{\rm SM}(0)$ are reference amplitudes used for normalization. Figure~\ref{fig:oofg}c shows the comparison between the OoF-corrected energies of the IF laser pulses and the Source Monitor energies. Their ratio, shown in Fig.\ref{fig:oofg}d, is stable on the long term of the run to $\pm 1\times 10^{-4}$.

\subsubsection{In-Fill Gain correction}
\label{sssec:in-fill}

After properly applying the OoFG correction, the laser calibration system becomes sensitive to the IFG effects.  In order to monitor and model the IFG function, the laser system interlaces laser pulses on a prescaled subset of the muon fills.  A pre-defined number of pulses are delivered at the prescaled rate and shifted by a prescribed amount on each subsequent prescaled fill.  By tuning the settings appropriately, the system samples the gain at fill times spanning from muon injection time to 
600~$\mu$s after injection.

\begin{figure}[htbp]
\centering
\includegraphics[width=0.8\linewidth]{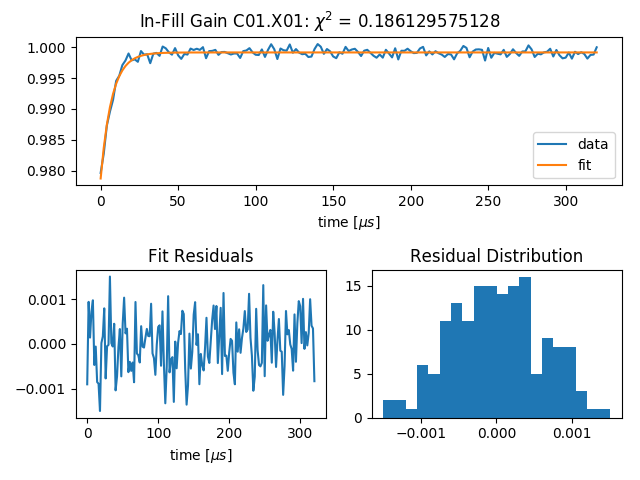}
\caption{An example of the measured and modeled IFG function. The data are fit with an exponential decay which returns to unity at long times. The residuals show that we can model the IFG to a precision of $\pm 4\times 10^{-4}$.}
\label{fig:ifgf}
\end{figure}

The IFG function is adequately modeled using an exponential decay returning asymptotically to unity as shown in Fig.~\ref{fig:ifgf}.  All individual crystal/SiPM pairs are modeled using in-fill laser data, and the positron events are then corrected using the model for the IFG.

\subsubsection{Short-Term Double-Pulse correction}
\label{sssec:dp}

The STDP correction cannot be modeled from standard positron data.  The laser system is prepared to operate in a secondary running mode to explore the STDP gain effects. This has been described in preceding Sec.~\ref{sssec:stdp}. These runs are taken with time separations from about 0 to 80~ns, and different amplitudes based on adjusting the position of wheels filled with neutral density filters within the laser optics.

\begin{figure}[htbp]
\centering
\includegraphics[width=0.75\linewidth]{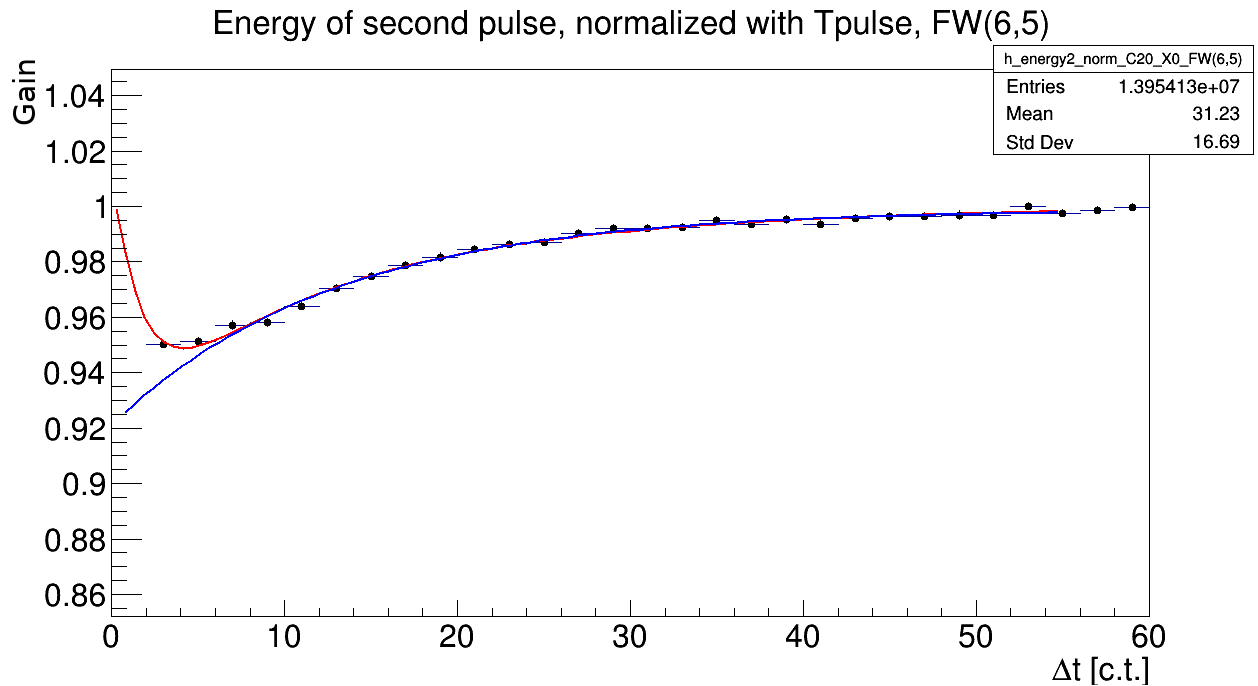}
\caption{Gain recovery response after STDP of a  SiPM for a single pair of filter wheel settings. Data are fit with both the exponential (blue) and the lognormal (red) functions (a clock tick corresponds to 1.25 ns).}
\label{fig:STDP:exp_lognorm}
\end{figure}

The  STDP gain shape has been measured in  
dedicated set of laser runs
in which the  delay
between the pulses 
varied from 0 to 60 clock ticks, in steps of 2 ticks (1.25 ns/tick). For each
delay, a few thousand laser events were collected over a time period of 5
minutes. The whole sequence was repeated for 18 combinations of pulse
heights, or energy deposits $E_1$ and $E_2$, using the filter wheels of both lasers.
The full procedure takes 3-4 days and it is repeated before the beginning of each data taking period, i.e. once per year.
During standard data taking, a one-hour-long sequence of Double Pulse measurements, with only one setting of the filter wheels, is taken  approximately once per week to verify the stability of the correction. Up to now, no effect has been observed at the required level of 
precision, but a detailed analysis with the full final statistics will eventually allow to correct for second order effects.
% setting the filter wheel of the first pulse in the 6 positions (5-10) and the one in front of laser 2 in the 3 positions (5, 7, 9).

The goal of the analysis is to find the energy dependence of the
parameters $a$ and $\tau$ of the gain shape defined as: 
\begin{equation}
G(\Delta t; i, E_1, E_2) = 1 - a(i,E_1,E_2)\cdot e^{-\frac{\Delta t}{\tau(i,E_1,E_2)}}
\label{eq:STDP:exp}
\end{equation}
where $i$ represents the SiPM number. In addition, another functional
form that describe a smooth return to $G=1$ at $\Delta t = 0$ is named
\textit{lognormal} and defined as: 
\begin{equation}
G(\Delta t) = 1 - a \cdot e^{-\frac{1}{2}\left[\log^2(\Delta
    t/\tau)\right]/\left[\log^2(t_M /\tau)\right]} 
\label{eq:STDP:lognorm}
\end{equation}
where $t_M$ and $a$ are the position and the amplitude of the minimum
and $\tau$ is the recovery time.

The analysis has shown that the time parameter $\tau$ does not depend
on the positron energy; this is expected, as the time constants depend
on the electronics.
On the contrary, the amplitude $a$, for both parametrizations, depends 
linearly on the energy
$E_1$  of the first positron with a slope that, on average, is equal
to $4.6\%/$GeV and $3\%/$GeV for the exponential and lognormal fit,
respectively\footnote{The difference is due to the fact that the
  parameter $a$ is extrapolated to $\Delta t=0$, for the exponential
  function, while it is the minimum function value,
  which normally happens at $4-5$ clock ticks, 
  for the lognormal distribution}.

 An example of a fit with the two functions is shown
in Fig.~\ref{fig:STDP:exp_lognorm}. 
Above $\Delta t = 6$ clock ticks they overlap, showing a sizable
difference only below $2$ clock tiks (note that each time interval is 2
clock ticks wide, with the central point in positions $\Delta t = 1,
3, 5, ...$ clock ticks). 

The behavior of the SiPM for very
short $\Delta t$ is not known and it would require specific studies;
however, pulses that are closer than 2 clock ticks ($2.5\,$ns)  
are merged by the pulse fitting algorithm and belong to another
category of events which is called {\em pile-up} and is studied
separately.  
Thus there is no practical difference between the two parametrizations, and the
exponential function is used for convenience.

Given the muon flux, the fraction of positrons which falls into the STDP 
category is about 3\% at $t=10$ microseconds after the muon injection, and it decreases as the muon flux
decays away. The average correction is about 1\%, for a flat muon 
distribution in the 80 ns range. These numbers correspond to a small but 
non-negligible effect of $3\times 10^{-4}$ between the beginning and the end of fill, where the rate is so
low that there is no short-term double-pulse effect.
This correction must be known at the $10^{-4}$ level or better, and thus a knowledge of 
the STDP parameters at the 10-20\% level is sufficient.

%\begin{figure}[!ht]
%	\centering
%	\includegraphics[width=\linewidth]{Exp_Lognorm.png}
%	\caption{Gain recovery of one SiPM. The blue curve is the exponential function, while the red curve is the Lognormal function.}
%	\label{fig:STDP:exp_lognorm}
%\end{figure}

%\section{LTDP JINST}
\subsubsection{Long-Term Double-Pulse correction}
\label{sssec:dp_long}

\begin{figure}[htbp]
	\centering
        \includegraphics[width=0.75\textwidth]{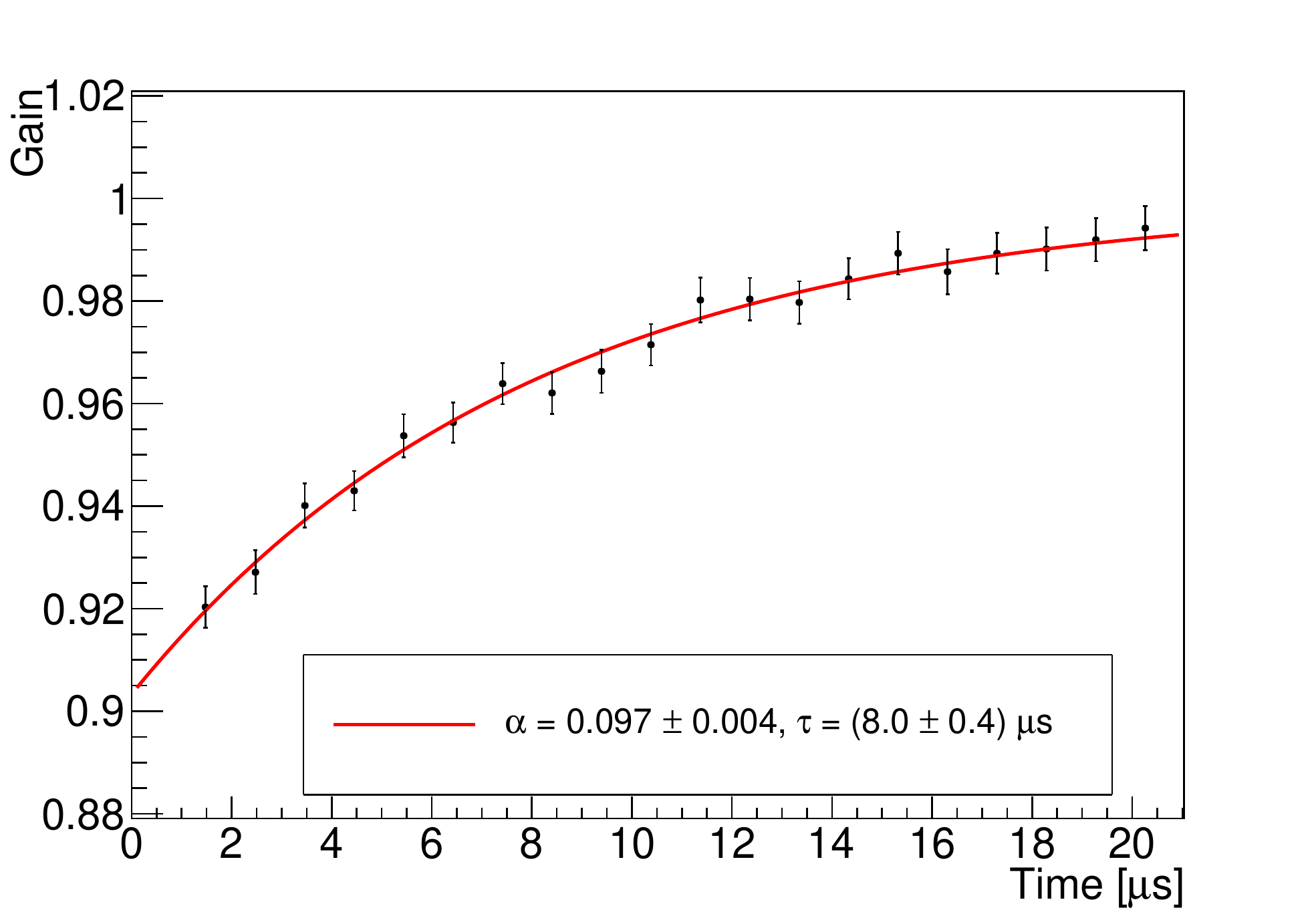}	
	\caption{The mean LTDP response for a calorimeter, for a given pair of filter wheel positions and 100 pulses in the burst. The fit is performed with an exponential function 1-$\alpha$e$^{-t/\tau}$. 
\label{fig:ltdp_response}}
\end{figure}

The LTDP study, as the STDP one, is an offline procedure, which uses data taken according to the Long-Time Double-Pulse operation mode described in Sec.~\ref{sssec:ltdp}. The aim of this study is to parametrize the gain function at the microsecond time scale, by mimicking the initial splash of muons and positrons that impinge on the calorimeters producing an overload of the HV power supply. This study is performed by varying the number of pulses in a burst and the energy of each pulse, by changing the filter wheel position. The probe, a second pulse after the burst, measures the gain value at a given time after the splash. By iterating this procedure and delaying the probe (in $1\, \mu$s step) from the last pulse in the burst we can reconstruct the gain function shown in Fig.~\ref{fig:ltdp_response}.
Data points are fitted with the function 1-$\alpha$e$^{-t/\tau}$, where
$\alpha$ quantifies the gain drop due to the burst energy and $\tau$ is the recovery time of the power supply. 
%The LTDP, unlike the STDP doesn't give additional corrections to OoFG and IFG as the 
%STDP, but it's 
The LTDP, unlike the STDP, does not give
additional corrections to the OoFG and IFG; it is however a powerful independent method to  characterize the  
behavior of each SiPM and obtain the gain sagging due to the recovery time of the 
SiPM bias voltage.

\section{Conclusions} 
The Muon $g-2$ laser calibration system has been designed with the primary goal of monitoring the gain fluctuations of the calorimeter photo-detectors at 0.04\% precision on short term scales (700~$\mu$s). Stability  below the per mille level is attained on longer time scales.
The adopted solution is based on a triggerable diode laser system with multiple laser heads with fluctuations below the percent level, an optical distribution system ensuring adequate intensity and homogeneity, and a system for monitoring the laser
system itself with a stability at the 0.01\% level.
Different operation modes have been described: the standard one, the double-pulse, the gain calibration and the flight simulator modes. All these modes are possible thanks to the choice of the optical components and the presence of the Laser Control Board which manages the interface between the experiment and the laser source, allowing the generation of light pulses according to specific needs. 
The performance of the laser system with data has been reviewed showing that it was able to meet the requirements of the experiment.

%BeginExpansion

%\section{Acknowledgement}
\section*{Acknowledgments}
Fruitful discussions with Fabrizio Scuri in the early stage of the experiment, and critical reviewing of the paper by Tim Gorringe are gratefully acknowledged.

This research was supported by Istituto Nazionale di Fisica Nucleare (Italy), by Fermi Research Alliance, LLC under Contract No. DE-AC02-07CH11359 with the United States Department of Energy, and by the EU Horizon 2020 Research and Innovation Program under the Marie Sklodowska-Curie Grant Agreement No.690835 and No.734303. This work was also partially supported by MSE grant No. 533-19-15-0022 and some equipment was funded by the EU Regional Development Fund project RC.2.2.06.-0001: Research Infrastructure for Campus-based Laboratories at the University of Rijeka.

%\section*{References}

%\bibliographystyle{unsrt}
%\bibliography{gminus2}

%\newpage

\end{document}